\documentclass{emulateapj}
\usepackage{epsfig,natbib,amsmath}

\newcommand{\mic}{\,$\mu$m }
\newcommand{\micpa}{\,$\mu$m}          
\newcommand{\muJy}{\,$\mu$Jy }
\newcommand{\muJypa}{\,$\mu$Jy}                              
\newcommand{\spi}{{\it Spitzer}}
\newcommand{\Lsol}{L$_\odot$}

\newcommand{\Lir}{L$_{\rm IR}$}

\bibpunct[]{(}{)}{;}{a}{}{,}
 
\shorttitle{The deep Spitzer 24micron observations of the COSMOS field}
\shortauthors{E.\,Le Floc'h et al.}

\begin{document}
\def\gtapp
{\mathrel{\hbox{\raise0.3ex\hbox{$>$}\kern-0.8em\lower0.8ex\hbox{$\sim$}}}}
\def\ltapp
{\mathrel{\hbox{\raise0.3ex\hbox{$<$}\kern-0.75em\lower0.8ex\hbox{$\sim$}}}}
\def\ts{\thinspace}

\title{Deep Spitzer 24\mic COSMOS imaging -- I. The evolution of  luminous dusty galaxies - Confronting the models}

\slugcomment{Accepted to The Astrophysical Journal,
22 July 2009}

\author{Emeric~Le~Floc'h\altaffilmark{1,2},
Herv\'e~Aussel\altaffilmark{3},
Olivier~Ilbert\altaffilmark{1,4},
Laurie~Riguccini\altaffilmark{1,3},
David~T.~Frayer\altaffilmark{5},
Mara~Salvato\altaffilmark{6},
Jason~Surace\altaffilmark{7},
Stephane~Arnouts\altaffilmark{8},
Chiara~Feruglio\altaffilmark{3},
Giulia~Rodighiero\altaffilmark{9},
Peter~Capak\altaffilmark{6,7},
Jeyhan~Kartaltepe\altaffilmark{1},
Sebastien~Heinis\altaffilmark{10},
Kartik~Sheth\altaffilmark{7},
Lin~Yan\altaffilmark{7},
Henry~Joy~McCracken\altaffilmark{11},
David~Thompson\altaffilmark{12},
David~Sanders\altaffilmark{1},
Nick~Scoville\altaffilmark{6},
Anton~Koekemoer\altaffilmark{13}
}

\altaffiltext{1}{Institute for Astronomy, University of Hawaii, 2680 Woodlawn Drive, Honolulu, HI 96822, USA}
\altaffiltext{2}{Spitzer fellow}
\altaffiltext{3}{UMR AIM (CEA--UP7--CNRS), CEA-Saclay, Orme des Merisiers, b\^at. 709, F-91191 Gif-sur-Yvette Cedex, France}
\altaffiltext{4}{Laboratoire d'Astrophysique de Marseille, BP 8, Traverse du Siphon, 13376 Marseille Cedex 12, France}
\altaffiltext{5}{Infrared Processing and Analysis Center, California Institute of Technology 100-22, Pasadena, CA 91125, USA}
\altaffiltext{6}{California Institute of Technology, MC 105-24, 1200, East California Boulevard, Pasadena, CA 91125, USA}
\altaffiltext{7}{Spitzer Science Center, California Institute of Technology, Pasadena, CA 91125, USA}
\altaffiltext{8}{Canada France Hawaii telescope corporation, 65-1238 Mamalahoa Hwy, Kamuela, Hawaii 96743, USA}
\altaffiltext{9}{Dipartimento di Astronomia, Universit\`a di Padova, Vicolo Osservatorio 2, I-35122, Padova, Italy}
\altaffiltext{10}{Johns Hopkins University, Bloomber Center \#380, 3400 N. Charles Street, Baltimore MD 21218, USA}
\altaffiltext{11}{Institut d'Astrophysique de Paris, UMR7095 CNRS, Universit\'e Pierre et Marie Curie, 98 bis Boulevard Arago, 75014 Paris, France}
\altaffiltext{12}{LBT Observatory, University of Arizona, 933 N. Cherry Ave., Tucson, Arizona, 85721-0065, USA}
\altaffiltext{13}{Space Telescope Science Institute, 3700 San Martin Drive, Baltimore MD 21218, USA}

\email{elefloch@ifa.hawaii.edu}

\begin{abstract} We present the first results obtained from the identification of $\sim$\,30\,000~sources in the {\it Spitzer}/24\mic observations of the COSMOS field at $S_{\rm 24\mu m}$\,$\gtapp$\,80\muJypa.  Using accurate photometric redshifts ($\sigma_z$\,$\sim$\,0.12 at $z$\,$\sim$\,2 for 24\mic sources with $i^+$\,$\ltapp$\,25\,mag~AB) and simple extrapolations of the number counts at faint fluxes we resolve with unprecedented detail the build-up of the mid-infrared background across cosmic ages. We find that $\sim$\,50\% and $\sim$\,80\% of the 24\mic background intensity originate from galaxies at $z$\,$\ltapp$\,1 and $z$\,$\ltapp$\,2 respectively, supporting the scenario where highly obscured sources at very high redshifts ($z$\,$\gtapp$\,2) contribute only marginally to the Cosmic Infrared Background.  Assuming flux-limited selections at optical wavelengths, we also find that the fraction of $i^+$--band sources with 24\mic detection strongly increases up to $z$\,$\sim$\,2 as a consequence of the rapid evolution that star-forming galaxies have undergone with lookback time.  Nonetheless this rising trend shows a clear break at $z$\,$\sim$\,1.3, probably due to $k$-correction effects implied by the complexity of spectral energy distributions in the mid-infrared. Finally, we compare our results with the predictions from different models of galaxy formation.  We note that semi-analytical formalisms currently fail to reproduce the redshift distributions observed at 24\micpa. Furthermore the simulated galaxies at $S_{\rm 24\mu m}$\,$>$\,80\muJy exhibit $R-K$ colors much bluer than observed and the predicted $K$--band fluxes are systematically underestimated at $z$\,$\gtapp$\,0.5.  Unless these discrepancies mainly result from an incorrect treatment of extinction in the models they may reflect an underestimate of the predicted density of high redshift massive sources with strong on-going star formation, which would point to more fundamental processes and/or parameters (e.g., Initial Mass Function, critical density to form stars, feedback, ...) that are still not fully controlled in the simulations.  The most recent backward evolution scenarios reproduce reasonably well the flux/redshift distribution of 24\mic sources up to $z$\,$\sim$\,3, although none of them is able to exactly match our results at all redshifts.  \end{abstract}


\section{Introduction}

Over the last decade, extragalactic infrared (IR) surveys have become a key component in our quest to understand galaxy formation.  The reprocessing of stellar and nuclear radiations by dust and the resulting emission of this energy at thermal IR wavelengths (8\micpa\,$\ltapp$\,$\lambda$\,$\ltapp$\,1000\micpa) plays a major role in shaping the panchromatic appearance of star-forming galaxies and active galactic nuclei \citep[e.g.,][]{Silva98,Gordon00,Nenkova02,Dopita05,Marshall07,Siebenmorgen07,daCunha08}. The direct characterization of the properties of these sources at IR wavelengths is therefore crucial for reliably estimating the fraction of their bolometric energy output absorbed by dust and converted into lower energy photons.  It also provides unique constraints on the nature of the different physical processes driving their on-going activity
 \citep[e.g,][]{deGrijp85,Mazzarella91,Genzel98,Dale06,Smith07,Spoon07}.
 
 The discovery of a large number of distant sources radiating copious amounts of energy in the IR \citep[e.g.,][]{Smail97,Barger98,Aussel99,Elbaz99,Blain02,Papovich04,Dole04,Daddi05} revealed that critical phases of the cosmic star formation and nuclear accretion history happened within dust-embedded environments hidden behind large amounts of absorbing material.  A significant fraction of the growth of structures over cosmic ages thus occurred in galaxies affected by substantial reddening and which characterization at shorter wavelengths (e.g., UV, optical) can be severely biased if their extinction is not properly estimated.  At large cosmological distances, highly obscured sources can even be barely detectable below $\sim$\,1\mic and the observations in the IR or the millimeter are sometimes the only way to unveil the presence of on-going star formation and/or nuclear activity in these galaxies \citep[e.g.,][]{Hughes98,Downes99,Houck05}.

 The infrared spectral energy distributions (SED) of star-forming galaxies and active galactic nuclei (AGN) usually peak at $\sim$\,80--200\micpa. The most reliable estimate of the energy absorbed by dust in these sources would thus require observations in the far-IR or the submillimeter.  Unfortunately though, data taken at such wavelengths have been so far hampered by source confusion as well as the very modest sensitivity of current detectors and the difficulty to access this frequency range from the ground. For instance, observations of the distant Universe performed at 70\mic and 160\mic with the {\it Spitzer Space Telescope} \citep{Werner04} or in the submillimeter with ground-based cameras of bolometers have been so far limited to the identification of
 galaxies at the very bright end of the luminosity function (i.e., L$_{\rm bol}$\,$\gtapp$\,10$^{12}$\,\Lsol),
 with biases due to dust temperature selection effects \citep[e.g.,][]{Blain04,Frayer06,Pope06,Borys06,LeFloch07}. Detecting the more typical high redshift sources at the peak of their IR SED has not been achieved yet, and our knowledge of the far-IR luminosity function of galaxies in the distant Universe is still affected by substantial uncertainties.

 This strong limitation can be partially remedied with observations performed at mid-IR wavelengths (i.e., 5\micpa\,$\ltapp$\,$\lambda$\,$\ltapp$\,60\micpa). Although cameras operating in the mid-IR are not as efficient as optical detectors they have been far more sensitive than in the far-IR, allowing the probe of the IR luminosity function down to much fainter levels. The observed correlations between the mid-IR and the far-IR luminosities of star-forming galaxies allow reasonable estimates of their total IR luminosity (\Lir\,=\,L$_{\rm 8-1000\mu m}$) with extrapolations from fluxes measured in the mid-IR \citep{Spinoglio95,Chary01,Bavouzet08,Symeonidis08}. At such wavelengths, starbursts and AGNs exhibit also quite disctinct spectral signatures, providing invaluable clues to quantify the relative contribution of star-forming and nuclear activity occurring within individual sources.

 Observations of the deep Universe with the 24\mic bandpass of the {\it Multi-band Imaging and Photometer for Spitzer} \citep[MIPS,][]{Rieke04} and with the {\it Infrared Spectrograph} \citep[IRS,][]{Houck04a} have led to tremendous progress in our understanding of the bolometric properties of distant sources \citep[e.g.,][]{Huang05,Yan05,Daddi05,Papovich06,Webb06,Reddy06,Menendez07,Sajina07,Lutz08,Pope08}. The \spi \, data provide tight constraints on the different mechanisms (merging, feedback, environmental effects) that regulate the activity of star formation and nuclear accretion in galaxies \citep[e.g.,][]{Noeske07,Elbaz07,Bridge07,Buat08,Bai07,Marcillac08}.  They also enable the identification of a large number of obscured AGNs that must have played a key role in driving the coeval growth of bulges and super-massive black holes but whose contribution is usually underestimated by surveys performed at other wavelengths \citep[e.g.,][]{Houck05,Polletta06,Alonso06,Dey08,Fiore08}.

 In spite of this wealth of results recently achieved with \spi \, at mid-IR wavelengths, a coherent and comprehensive picture of the deep IR Universe is still missing though. Given the variety of starburst and AGN SEDs in the mid-IR \citep[e.g.,][]{Laurent00,Brandl06,Weedman05,Wu06,Armus07}, observations at these wavelengths are affected by quite complex $k$-correction effects and their sensitivity decreases rapidly beyond $z$\,$\sim$\,1. Hence, wide and shallow surveys are only sensitive to the very bright end of the mid-IR luminosity function, while the deeper observations usually performed at the expense of the covered area suffer from limited statistics.  Also, the identification of mid-IR selected sources at high redshift can be extremely difficult given their faintness at optical wavelengths.  As a result the exact role that dusty galaxies played with respect to the less luminous but much more numerous sources at the faint end of the luminosity function is still under debate \citep[but see][]{Reddy08}.
 
 In the attempt to unify the different pictures of galaxy evolution recently achieved by optical and IR surveys
 we undertook deep 24\mic imaging over the whole area of
 the COSMOS field
 using the MIPS instrument on-board \spi. The {\it Cosmic Evolution Survey\,} (COSMOS) is designed to probe the evolution of galaxies and AGNs up to $z$\,$\sim$\,6 over a sky region large enough to address the role of environment and large scale structures \citep{Scoville07a}. It is based on deep multi-wavelength observations
 performed over $\sim$\,2\,deg$^2$ from the X-ray to radio wavelengths \citep{Hasinger07,Taniguchi07,Sanders07,Bertoldi07,Schinnerer07}, with additional support from high spatial resolution imaging obtained with the {\it Hubble Space Telescope} \citep{Scoville07b,Koekemoer07}. Intensive spectroscopic follow-up was also carried out with the {\it Very Large Telescope} \citep{Lilly07} while photometric redshifts with unprecedented accuracy were determined for more than $\sim$\,900\,000 $i^+$--band selected galaxies in the field \citep{Ilbert09}.
 The quality of the data sets available in COSMOS (and in particular the depth with respect to the covered area)
 is therefore very well suited for probing the multi-wavelength properties of
 galaxies with minimized cosmic variance and over a wide range of redshifts and luminosities.

 In this paper we present our very first analysis of the mid-IR selected galaxy population detected down to $S_{\rm 24\mu m}$\,$\sim$\,80\muJy in COSMOS. Data are described in Sect.\,2 and our 24\mic source number counts are shown in Sect.\,3.  Using the photometric redshifts of \citet{Ilbert09} and \citet{Salvato09} we then determine the 24\mic redshift distributions in Sect.\,4 and we analyze the number counts as a function of cosmic time in Sect.\,5 to constrain the history and the build-up of the Cosmic Infrared Background (CIB). In Sect.\,6 we present comparisons between the observed 24\mic source redshift distributions and the predictions from both phenomenological and semi-analytical scenarios of galaxy formation. The results are discussed in Sect.\,7, which allows us to quantify how well the contribution of dusty luminous sources as a general population is currently understood in models of galaxy evolution. Future papers will be dedicated to the analysis of the multi-wavelength properties of the COSMOS mid-IR selected galaxies, their evolution, their clustering and their connection with sources selected with other criteria.  Throughout this paper we assume a $\Lambda$CDM cosmology with H$_0$\,=\,70~km~s$^{-1}$\,Mpc$^{-1}$, $\Omega_m$\,=\,0.3 and $\Omega_{\lambda}\,=\,0.7$ \citep{Spergel03}.  Magnitudes are quoted in AB\footnote{mag (AB) = --2.5\,$\times$\,log$_{\rm 10}$ [F$_{\nu}$\,(mJy)]\,+16.4.} unless the use of the Vega system is explicitly stated (e.g., Sect.\,\ref{sec:compare_lacey}).

\section{The data}

\subsection{The 24\mic observations of COSMOS}

\subsubsection{Data acquisition and reduction}

The COSMOS field was observed at 24\mic with the MIPS instrument on-board \spi \, as part of two General Observer programs (PI D.\,Sanders).  The first observations (GO2, PID 20070) were carried out in January 2006 with the MIPS medium scan mode and two scan passes per sky pixel. As described in more detail by \citet{Sanders07} this shallow imaging covered a total area of $\sim$\,4\,deg$^2$ centered on COSMOS, with a median integration time of 80\,s per pixel.  This first program was also used to test in a small region of the field the feasibility to improve the 24\mic sensitivity with longer integrations despite the COSMOS mid-IR background being $\sim$\,2$\times$ higher than in other typical ``cosmological'' fields (e.g., {\it Hubble/Chandra Deep Fields}, Lockman Hole).  After validating this additional step, deeper imaging was performed in 2007 over the whole COSMOS field as part of a second program using the MIPS slow scan mode (GO3, PID 30143).  These data yielded a second and independent 24\mic coverage over a total area of $\gtapp$\,3\,deg$^2$. Across the nominal 2\,deg$^2$ of COSMOS the combination of the GO2 and GO3 observations results in a median integration time of $\sim$\,3\,360\,s per sky pixel. Note that the MIPS scan mode provides simultaneous imaging at 24, 70 and 160\micpa; the observations of the COSMOS field at 70/160\mic are presented by Frayer et al. (submitted to AJ).

All the data set was reduced following the procedure outlined in \citet{Sanders07}. For each observing campaign a background pattern derived for every scan mirror position of the instrument was scaled and subtracted from all individual frames. These background-subtracted images were then interpolated to a common grid of pixels using the MOPEX package \citep{Makovoz05}. They were finally coadded with our own IDL routines using the median of all the pixels associated with each sky position.  We stress that the use of a median combination was necessary in order to reject moving sources present in the field at the time of the observations. A forthcoming paper (Aussel et al. in prep.) will describe in more detail the data reduction. It will also provide a list of the 24\micpa--detected asteroids in COSMOS.

\subsubsection{24\mic source extraction, photometry}

\begin{figure*}[htpb]
  \epsscale{1.1}
  \plotone{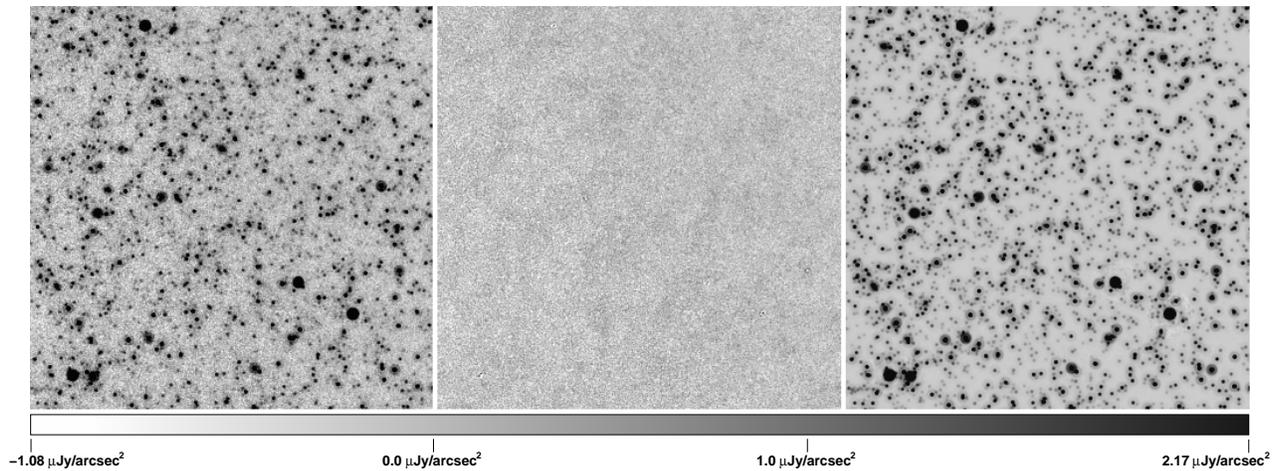}
 \caption{A 15'$\times$15' sub-region of the COSMOS field observed at 24\mic (left panel), along with the residual image obtained on the same field of view after subtracting  sources with a PSF fitting technique (central panel). The  ``reconstructed" image derived with this PSF fitting is shown on the right panel. }
\label{fig:daophot}
\end{figure*}

The 24\mic sources in our mosaic were detected using the automatic procedure of the Sextractor software \citep{Bertin96} and we measured their flux densities with the Point Spread Function (PSF) fitting technique of the DAOPHOT package \citep{Stetson87}.  Given the Full-Width at Half Maximum (FWHM) of the MIPS-24\mic PSF ($\sim$\,6\arcsec), most of the sources are not resolved in our data.  In deep and crowded images such as the ones typically obtained in the MIPS deep surveys, the PSF fitting thus provides more reliable flux measurements than aperture photometry. Furthermore, DAOPHOT performs simultaneous fits to multiple objects, which ensures a proper separation of blended sources especially when the fluxes are close to the confusion limit. No prior information from any other wavelength was used for the source detection itself. However, several PSF fitting iterations were necessary to achieve a clean extraction of the blended sources that were not properly identified by Sextractor in the input 24\mic image. In brief, these blended cases were located (with Sextractor again) as positive detections in the residual mosaic obtained after the first PSF fitting. Their positions were added to the initial list of 24\mic sources determined with Sextractor and another extraction of the initial COSMOS 24\mic image was then performed. Three iterations of this process were carried out, until the residual mosaic was cleaned from any positive signal above the detection threshold chosen for Sextractor.  The result of this PSF fitting is illustrated in Figure~\ref{fig:daophot}. It shows a sub-region of our COSMOS 24\mic mosaic along with the corresponding residual image obtained after subtracting all sources extracted with DAOPHOT.

To run this PSF fitting we built an empirical PSF from the brightest point sources identified in the mosaic and we subsequently fit this PSF to every single object detected by Sextractor. DAOPHOT performs this fit by scaling the empirical PSF within a fixed circular aperture centered on the source and it provides the flux of the fitted PSF {\it enclosed within this aperture}. Therefore a correction must be applied to account for the spatially-extended wings of the PSF lying beyond the radius of the chosen aperture.  This correction was determined using the MIPS-24\mic STiny Tim PSF models provided by the {\it Spitzer Science Center}. These models allow to account for the fraction of energy lying at large radii where the signal to noise in our data is not large enough to constrain the exact profile of the Point Spread Function.  Relying on such corrections we believe that our flux absolute calibration should be accurate within a few percent, which is consistent with the results obtained by \citet{Engelbracht07} and \citet{Rieke08} as part of the calibration of the MIPS 24\mic detector.

Following the convention adopted by the {\it Spitzer Science Center} we assumed a stellar 10\,000\,K black-body spectrum as the reference underlying SED for the 24\mic flux density measurements.  We tested this calibration by cross-correlating our source list with the stars of the {\it Two Micron All Sky Survey\,} (2MASS) catalog \citep{Jarrett00}. Our comparison did not reveal any systematic deviation from the relation $K_s - [24] =0$ (Vega, with [24] denoting the equivalent magnitude measured in the 24\mic MIPS bandpass), which is expected for most of the stellar spectral types when no dust emission excess is detected above the photospheric stellar emission \citep[e.g.,][]{Gorlova06}. This cross-correlation also allowed us to register the absolute astrometric calibration of our 24\mic mosaic to the 2MASS astrometric system. This was performed by applying a systematic offset of $\delta$\,Dec\,=\,$-$0.3\arcsec \, to the astrometry derived from the \spi \, data products.

The noise in the final 24\mic mosaic was characterized using two different methods. First, we computed the dispersion of a set of random flux measurements performed over blank field regions of the COSMOS 24\mic image within the same aperture as the one used for the PSF fitting.  Similar to the photometry measurements described above a flux correction was applied to account for the extent of the PSF lying outside of the aperture, which led to an equivalent sensitivity limit of 1$\sigma$\,=\,18\muJypa. Second, we also performed a number of simulations adding artificial point-like sources in blank field regions of the mosaic.  We measured the fluxes of these sources with the PSF fitting of DAOPHOT and the distribution of the differences between their measured and input fluxes was fitted with a gaussian function. Our best fit was obtained for a 1$\sigma$~standard deviation varying between 14\muJy and 16\muJy depending on the range of input fluxes. Using these simulations we finally examined the robustness of the error measurements provided by our PSF fitting by comparing the DAOPHOT uncertainties obtained for the simulated point sources with the errors directly estimated from the differences between their measured and input fluxes. We applied a systematic scaling of 3.8 to all the DAOPHOT errors so as to get 68\% of the distribution of these differences within the DAOPHOT 1$\sigma$ uncertainties.

The completeness of our 24\mic source extraction was determined using the same point source simulations as the ones described in the previous paragraph. We inserted within the mosaic a set of empirical PSFs that were scaled to provide a flux distribution representative of the one characterizing the 24\mic extragalactic source population. These PSFs were subsequently extracted following the same method as the one employed to detect and to measure the flux of the sources already present in the 24\mic image. We found that our source extraction is more than 90\% complete above a 24\mic flux $S_{\rm 24\mu m}$\,$\sim$\,80\muJypa, which should thus be considered as a safe and conservative flux limit for any analysis requiring an unbiased 24\mic selection of sources in the COSMOS field. According to the simulations though, our PSF fitting is still reliable down to fluxes as faint as $\sim$60\muJy and in spite of a lower completeness ($\sim$\,75\%) our source extraction should suffer from a relatively small contamination by fake objects at these flux levels.
 
Based on this characterization we defined our final COSMOS MIPS-24\mic catalog as the population of 24\mic sources brighter than 60\muJy lying within the area covered by the COSMOS photometric redshift catalog of \citet[, see section~\ref{sec:photoz}]{Ilbert09} but outside of the masked regions defined from the Subaru optical observations of the COSMOS field.  These masked areas cover a surface as large as 0.26\,deg$^2$ but the optical photometry in these regions is affected by larger uncertainties due to the presence of very bright or even saturated objects. In particular, the photometric redshifts in these areas are less accurate. Keeping these regions would thus compromise our analysis of the MIPS-selected galaxies given the large fraction of 24\mic sources associated with faint optical counterparts.  Our 24\mic catalog represents a total of 39\,413 sources covering an effective area of 1.68\,deg$^2$. In the remaining of this Section~2 we will describe the identification of the 24\mic source counterparts down to $S_{\rm 24\mu m}$\,$=$\,60\muJypa. This characterization will supplement other studies carried out by our team using selections performed at other wavelengths. However, the analysis reported in some other sections of this current paper assume an unbiased 24\mic selection and only sources above 80\muJy will be considered in this case.

\subsection{Identification of  24\mic source counterparts}
\label{sec:match}

To ensure the most reliable and most complete identification of the 24\mic source counterparts, we first correlated the 24\mic data with the $K_s$--band COSMOS catalog of McCracken et al. (submitted to ApJ).  As we will see later a non-negligible fraction of the 24\mic sources at $z$\,$\gtapp$\,1 are associated with extremely faint optical counterparts. A direct cross-correlation between the MIPS-24\mic and the optical catalogs of COSMOS would thus result in a lower rate of identifications. Furthermore, the larger uncertainties affecting the determination of source centroids at faint 24\mic fluxes, the width of the 24\mic PSF and the very high density of optical sources in our data (e.g., $\sim$\,45 galaxies~arcmin$^{-2}$ for $i^+_{\rm AB}$\,$<$\,25\,mag in the Subaru COSMOS observations, \citealt{Taniguchi07, Capak07}) could lead to lots of wrong associations if these identifications had to rely on a direct match between the 24\mic and optical source catalogs. On the contrary, the COSMOS $K_s$--band catalog is deep enough ($K_{s}$\,$\sim$\,23.7\,mag at 5$\sigma$, McCracken et al.) to detect counterparts for the majority of the 24\mic selected sources down to our MIPS sensitivity limit. The density of $K_s$--band sources is also substantially smaller ($\sim$\,7 galaxies~arcmin$^{-2}$ for $K_{s}$\,$<$\,22\,mag) than in the optical bands and the near-IR wavelengths are ``closer" to the 24\mic bandpass, which minimizes the risk of wrong associations between mid-IR and $K_s$-selected sources.  Finally we note that these $K_s$--band observations of COSMOS were executed under excellent seeing conditions ($\sim$\,0.7\arcsec \, at 2.2\micpa). Compared to the other COSMOS near-IR data taken with e.g., the IRAC instrument (FWHM\,$\sim$\,1.6\arcsec \, at 3.6\micpa, \citealt{Sanders07}), the PSF in the $K_s$--band image is much narrower and allows more robust identifications in the case of sources blended at 24\micpa.

We correlated the 24\mic and $K_s$--band catalogs with a matching radius of 2\arcsec. We found that this distance was large enough to enable the identification of most of the 24\mic sources, taking into account the width of the 24\mic PSF (FWHM\,$\sim$\,6\arcsec) and the uncertainty on the astrometric position of each single 24\mic source. This radius was nonetheless sufficiently small to minimize the fraction of multiple matches as well as the association between objects randomly-aligned on the sky.  Among the 39\,413 MIPS--24\mic sources of the catalog, we identified a single $K_s$--band counterpart for 33\,146 objects (84\% of the sample) while
two possible matches were obtained for 4\,470 other 24\mic detections ($\sim$\,11.5\%).  For theses cases we kept the closest possible counterpart when its centroid was at least twice closer to the 24\mic source than the second possible match (2\,320 objects) while additional priors from the COSMOS IRAC-3.6\mic observations were considered to assign the most likely counterpart to the other 24\mic detections.  No association was found for the remaining 1\,797 sources (4.5\% of the catalog). A cross-correlation between these unidentified sources and our 3.6\mic catalog using the same matching radius revealed IRAC counterparts for 437 objects, a rare population that must be characterized by extremely steep SEDs between 2\mic and 24\micpa. For the other sources a visual inspection of our data showed that some of them may have counterparts either fainter than the detection thresholds of our $K_s$/IRAC COSMOS catalogs or with a centroid located slightly beyond our 2\arcsec \, matching radius. Many of the unidentified objects have also a signal/noise below 5$\sigma$ at 24\micpa. At the faintest 24\mic flux leves some could originate from a small contamination of our catalog by fake sources. It should however not significantly bias our subsequent results given the small fraction that they represent among the whole sample of 24\mic sources.  In summary we identified secure $K_s$--band counterparts for $\sim$\,90\% of the COSMOS 24\mic catalog down to 60\muJypa, while less robust identifications were still found for another 5.5\% of the MIPS-24\mic source population.

Finally we note that this cross-identification with near-IR data also revealed 53 pairs of 24\mic sources associated with the same $K_s$--band counterpart.  A careful inspection of the 24\mic mosaic revealed that they correspond to single objects incorrectly deblended by the PSF fitting performed in the MIPS data. These ``double'' components were merged together to form a unique 24\mic source, reducing the MIPS-selected catalog to 39\,360 objects.

\vspace{.3cm}

\subsection{Photometric redshifts}
\label{sec:photoz}
\subsubsection{Input catalogs}

The distance of the 24\mic sources was characterized using the photometric redshifts derived by \citet{Ilbert09} and \citet{Salvato09} for the optically and X-ray selected sources of the COSMOS field.  These photometric redshifts were determined with deep photometry (e.g., Capak et al. 2009, in preparation) performed in a total of 30~large, medium and narrow band filters over a wavelength range covering the far-UV at 1550\AA \, up to the mid-IR at 8.0\micpa. This large amount of data and the high quality of the observations used in their work resulted in a substantial improvement of photometric redshift accuracy with respect to the determination of redshifts performed in other fields.

The catalog of \citet[, hereafter I09]{Ilbert09} provides redshifts for 933\,789 sources selected at $i^+_{\rm AB}$\,$<$\,26.5\,mag from the Subaru/Suprime--CAM observations of COSMOS, with a dispersion as small as $\sigma_{\Delta z/(1+z)} = 0.012$ for $i^+_{\rm AB}$\,$<$\,24\,mag and $z$\,$<$\,1.25.  This unprecedented accuracy in photometric redshift determinations was mostly achieved thanks to the high sampling of the observed spectral energy distribution of sources around the position of their Balmer break as well as a new method accounting for the contribution of emission lines to their measured flux densities.  As shown by \citet{Ilbert09} the uncertainties affecting their redshift estimates depend primarily on the redshift and the apparent $i^+$--band magnitude of sources (see their figure~9), with errors naturally increasing for fainter and more distant galaxies.  However a comparison with faint spectroscopic samples obtained in the COSMOS field (\citealt{Lilly07}) revealed a dispersion $\sigma_{\Delta z/(1+z)}$ as low as 0.06 for sources with 23\,mag\,$<$\,$i^+_{\rm AB}$\,$<$\,25\,mag at 1.5\,$\ltapp$\,$z$\,$\ltapp$\,3.  These small uncertainties will be critical for our analysis of the MIPS-selected galaxy population given the large fraction of 24\mic sources associated with faint (i.e., $i^+_{\rm AB}$\,$\gtapp$\,24\,mag) optical counterparts at $z$\,$\gtapp$\,1.

The redshifts provided by I09 were determined using a library of star-forming and passive galaxy templates and they are not adapted to galaxies dominated in their optical/near-IR continuum by the contribution of an active galactic nucleus (such as the type~1 AGNs and optically-luminous quasars). This issue must be treated with care given the non-negligible contribution of AGNs to IR-selected galaxy samples \citep[e.g.,][]{Genzel98,Houck05}.  The redshifts of the 24\mic sources also identified with an X-ray counterpart were therefore supplemented using the catalog of \citet{Salvato09}, which provides photometric redshifts for 1542 X-ray selected sources detected in the $XMM$-Newton observations of COSMOS \citep{Hasinger07,Brusa07}. \citet{Salvato09} used an optimized set of templates with hybrid combinations of AGN and non-active galaxy SEDs, which allowed them to derive photometric redshifts with an accuracy comparable to that achieved by \citet{Ilbert09}.  They also made an extensive use of multi-epoch observations of COSMOS to correct for the time-variability of type~1 AGNs. The photometric redshifts of the full $XMM$-selected sample mostly lie at 0.5\,$\ltapp$\,$z$\,$\ltapp$\,1.5 but extend up to $z$\,$\sim$\,3, with an accuracy as good as $\sigma_{\Delta z/(1+z)}=0.015$ for faint sources ($i^+_{\rm AB}$\,$\gtapp$\,22.5\,mag) at $z$\,$\sim$\,1 \citep{Salvato09}.

We stress that the redshifts of I09 were also used in the case of the 24\mic sources associated with obscured nuclei (such as sources characterized by a rising power-law continuum in the IRAC bands) but with no X-ray detection. Their emission at optical wavelengths is usually dominated by the contribution of their host galaxy and the photometric redshifts of I09 should not be significantly less accurate than those determined for star-forming sources. This argument is actually reinforced by the rather good agreement that we observed betweeen the redshifts of \citet{Salvato09} and those of \citet{Ilbert09} for the type~2 AGNs common to both samples.

\subsubsection{Photometric redshifts of 24\mic sources}

To derive the photometric redshifts of the 24\mic sources we first correlated the list of their $K_s$--band counterparts with the catalog of \citet{Ilbert09} using a matching radius of 1\arcsec. We found a single optical counterpart at $i^+_{\rm AB}$\,$<$\,26.5\,mag for 35\,347 $K_s$--band sources (90\% of the full 24\mic source catalog). Double matches were obtained for 361 other $K_s$--band objects. In those cases, we noticed that the difference between the matching distances of the first and the second closest matches was large enough to consider the first one as the most likely optical counterpart. The catalog of I09 is limited to $i^+_{\rm AB}$\,$=$\,26.5\,mag and no optical match could be found for the remaining 1\,855 $K_s$--band sources (i.e., $i^+_{\rm AB}$\,$>$\,26.5\,mag).

From this match between the $K_s$ and the optical bands we then identified 1\,129 optical counterparts also associated with a source detected by $XMM$. Among those, 1\,111 are included in the analysis presented by \citet{Salvato09} and we thus assigned to these objects the photometric redshifts provided in their catalog of X-ray sources. The identification of the optical counterparts associated with the remaining 18 $XMM$ objects is considered less secure \citep{Brusa07}.  For these objects we kept the original photometric redshift derived by I09.

Finally we considered the list of the 1\,797 MIPS-24\mic sources for which we did not find any $K_s$--band counterpart (see Sect.\,\ref{sec:match}). We correlated this list with the catalog of \citet{Ilbert09} using a matching radius of 2\arcsec, assuming that some of these sources could have counterparts detected in the very deep Subaru $i^+$--band observations of COSMOS \citep{Taniguchi07} but not reported in our $K_s$--band catalog.  Among these 24\mic sources, 667~objects were identified with one single counterpart in the optical catalog.  We note that some of them seem qto be trully detected in the $K_s$--band image of COSMOS (McCracken et al.), although their associated signal to noise is very low. As previously reported some of them also have counterparts at 3.6\mic in the IRAC data. These ``by-eye" detections gave us further confidence in the reliability of their identification in the $i^+$--band catalog of I09.

To summarize, we derived photometric redshifts for $\sim$\,92\% objects of the initial 24\mic catalog selected above 60\muJypa. Sources with counterparts in the $K_s$--band but no identification at $i^+_{\rm AB}$\,$<$\,26.5\,mag account for a fraction of 5\%, while 3\% of the catalog correspond to 24\mic sources with no identification in the optical nor in the near-IR.  If we rather consider a selection above 80\muJypa, we find a total of 29\,410 sources in the initial MIPS-24\mic catalog and 94\% of these objects are identified with a redshift. Independently of their possible association with a $K_s$--band or 3.6\mic counterpart, the 24\mic sources with no redshift are very faint at optical wavelengths. Previous studies have shown that they mostly correspond to highly-obscured luminous galaxies at 1\,$\ltapp$\,$z$\,$\ltapp$\,3 rather than low redshift sources populating the faint end of the luminosity function \citep[e.g.,][]{Houck05,Dey08}.

\subsubsection{Characterization of the photometric redshift uncertainties}

By construction an infrared-selected sample of sources is biased toward dusty galaxies. The determination of photometric redshifts for these objects is subject to more uncertainties if the effect of dust extinction is not properly taken into account.  Furthermore, a large fraction of mid-IR sources in the distant Universe are associated with very faint optical counterparts also characterized by less accurate redshifts.  The uncertainties affecting the photometric redshifts of the COSMOS MIPS-24\mic sources should thus be carefully assessed before analyzing our sample further.

\begin{figure}[htpb]
  \epsscale{1.1}
  \plotone{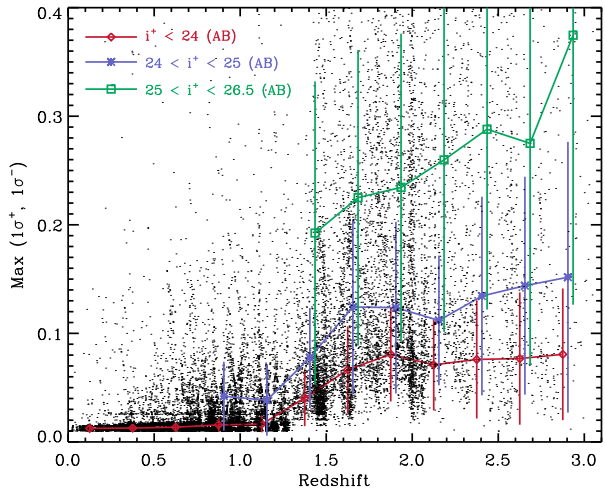}
 \caption{Photometric redshift uncertainties ($\sim$\,1\,$\sigma$) of the COSMOS 24\micpa--selected sources with S$_{\rm 24\mu m}$\,$>$\,80\muJy (dots), estimated using the probability distribution functions of \citet{Ilbert09} and represented as a function of photometric redshift. The solid lines represent the median values calculated in three bins of
   $i^+$--band magnitude (open diamonds: $i^+$$<$\,24\,mag; asterisks: 24\,mag\,$<$\,$i^+$$<$\,25\,mag; open squares: 25\,mag\,$<$\,$i^+$$<$\,26.5\,mag) and using
   at least 50 sources per bin of redshift. The associated error bars were computed from the median absolute deviation of each underlying distribution.  [{\it See the electronic edition of the Journal for a color version of this figure.}]}  \label{fig:z_uncert} \end{figure}

\citet{Ilbert09} presented a detailed comparison between their photometric redshifts and the spectroscopic redshifts of MIPS-24\mic sources determined by \citet{Lilly07} and Kartaltepe et al. (in prep.)  with spectroscopic follow-ups of the COSMOS field.  Up to $z$\,$\sim$\,1.5 I09 found median dispersions $\sigma_{\Delta z/(1+z)}$ as low as 0.01 and 0.05 for 24\mic sources with 22.5\,mag\,$<$\,$i^+_{\rm AB}$\,$<$\,24\,mag and 24\,mag\,$<$\,$i^+_{\rm AB}$\,$<$\,25\,mag respectively.  For mid-IR galaxies at higher redshifts and/or with fainter optical counterparts there is however no spectroscopic sample currently available in COSMOS to allow such comparison. To characterize the uncertainties of their photometric redshifts we relied on their associated Probability Distribution Functions PDF(z) as provided by I09. For their optically-selected sample \citet{Ilbert09} showed that the 1$\sigma$ uncertainties\footnote{Here we define the positive 1$\sigma^+$ and negative 1$\sigma^-$ photometric redshift uncertainties using the relation $\chi^2(z_{\rm best} + 1\sigma^+) = \chi^2(z_{\rm best} - 1\sigma^-) = min(\chi^2(z)) + 1$, with $z_{\rm best}$ the most likely photometric redshift estimated from the minimum of the merit function $\chi^2(z)$.  For each object the final 1$\sigma$ uncertainty is taken as the maximum between 1$\sigma^+$ and 1$\sigma^-$.} derived from these PDF(z) provide a reliable estimate of the dispersion $\sigma_{\Delta z/(1+z)}$ obtained from the comparison with spectroscopic redshifts. In Figure~\ref{fig:z_uncert} we show these 1$\sigma$ uncertainties restricted to the MIPS-24\mic sources as a function of their photometric redshifts, along with their median value calculated in three different bins of $i^+$--band magnitude. Up to the highest redshift bins characterizing our sample (i.e., $z$\,$\ltapp$\,3) we see that the photometric redshifts are still quite robust up to $i^+_{\rm AB}$\,$<$\,25\,mag (1$\sigma$\,$\ltapp$\,0.15). At fainter magnitudes, the uncertainties are somehow larger but the dispersions are still as low as 1$\sigma$\,$\ltapp$\,0.25 up to $z$\,$\sim$\,2 where most of the faint sources in our sample are located (see Sect.\,3).

\begin{figure}[htpb]
  \epsscale{1.1}
  \plotone{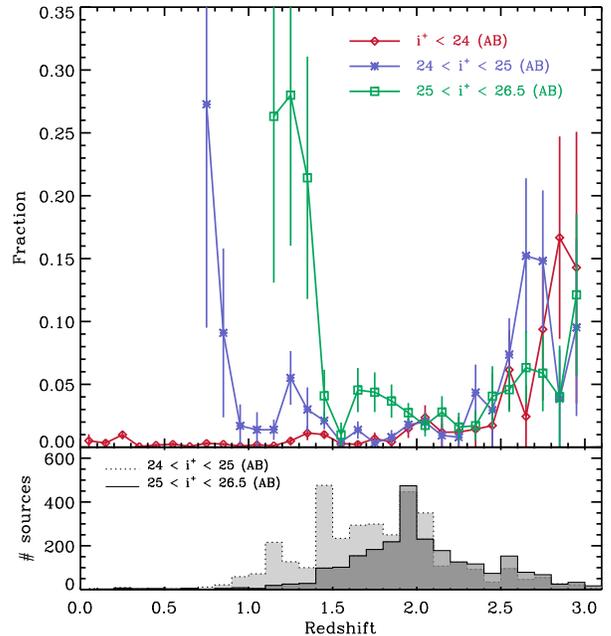}
  \caption{Fraction of 24\mic sources with S$_{\rm 24\mu m}$\,$>$\,80\muJy and characterized by a redshift probability distribution function with two different peaks, illustrated as a function of redshift for the same bins of $i^+$--band magnitude as those considered in Figure~\ref{fig:z_uncert} (top panel).  Each value was estimated using at least a total of 10~sources per bin of redshift and $i^+$--band magnitude.  Error bars were computed from Poissonian uncertainties.  The bottom panel shows the redshift distributions of the 24\mic sources associated with counterparts at 24\,mag\,$<$\,$i^+$$<$\,25\,mag (light grey shaded region) and 25\,mag\,$<$\,$i^+$$<$\,26.5\,mag (dark grey shaded region). Note the very small fraction of outliers up to $z$\,$\sim$\,2.5. The increasing trend observed at $i^+$\,$>$\,24\,mag and 0.8\,$\ltapp$\,$z$\,$\ltapp$\,1.3 affects only a relatively small number of sources.  [{\it See the electronic edition of the Journal for a color version of this figure.}]}
\label{fig:double_z}
\end{figure}

The larger uncertainties that may affect the photometric redshifts of faint optical sources can be produced by the presence of multiple peaks in their PDF(z). To quantify the contribution of these potential catastrophic failures we considered the fraction of MIPS-24\mic sources with a PDF($z$) showing a secondary local maximum stronger than 10\% of the value of its first peak. These fractions are illustrated as a function of redshift and for different bins of optical magnitudes in Figure~\ref{fig:double_z}. They show that up to $z$\,$\sim$\,2.5 the number of possible outliers is negligible ($\ltapp$\,5\%) for most of our sample.
Fractions reaching $\sim$\,25\% may be observed at $i^+_{\rm AB}$\,$>$\,24\,mag and 0.8\,$\ltapp$\,$z$\,$\ltapp$\,1.3, yet these estimates only affect a very small number of sources (see bottom panel of Figure~\ref{fig:double_z}).

\vspace{.5cm}

\section{Differential source number counts}
\label{sec:counts}

\begin{figure}[htpb]
  \epsscale{1.1}
  \plotone{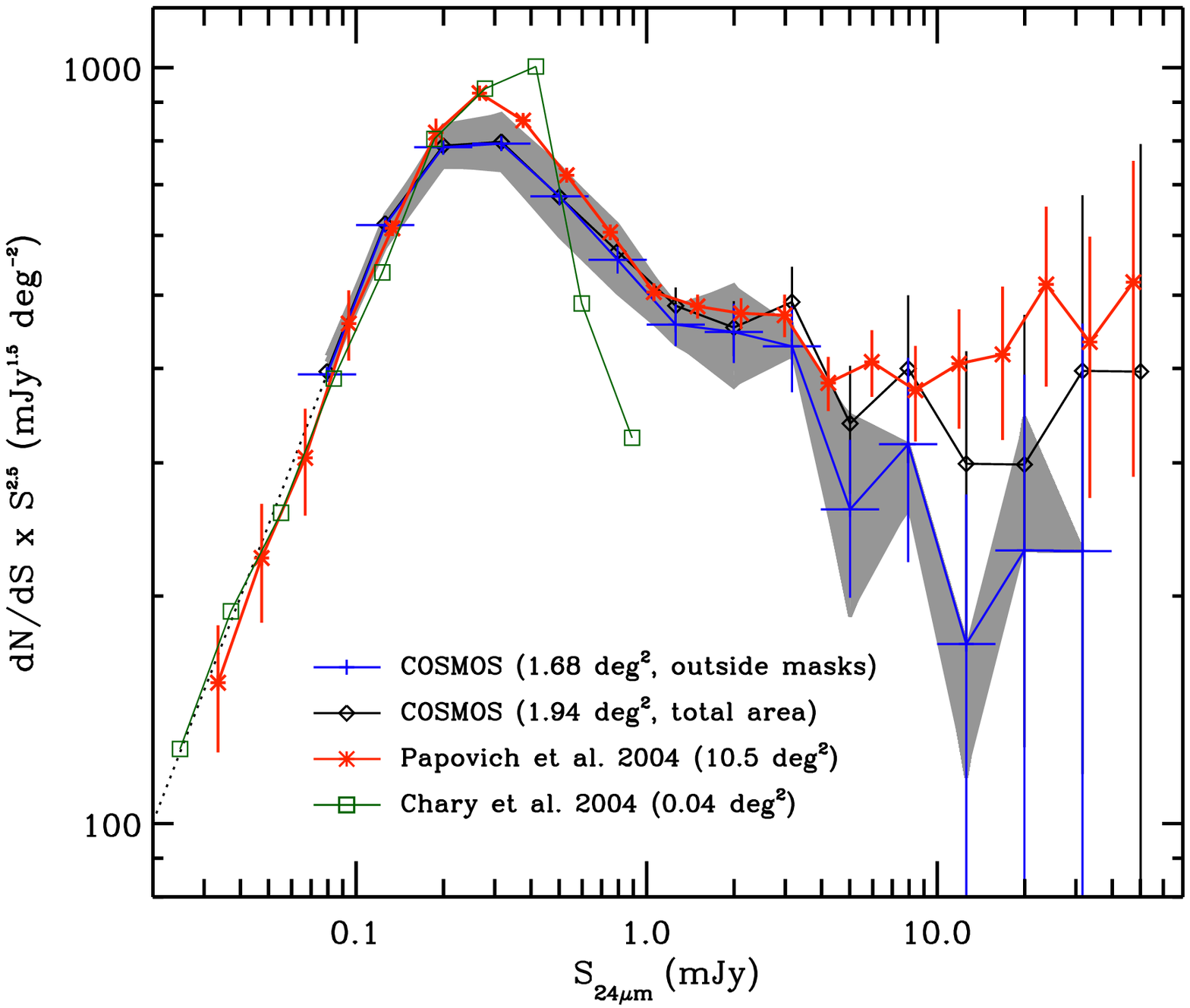}
 \caption{Differential 24\mic source number counts derived from the COSMOS field outside of the optically-masked regions ('plus' symbols). The grey shaded area 
 illustrates the effect of adding
 a systematic uncertainty of $\pm$\,5\% to the absolute 24\mic flux calibration.  The counts derived for the full coverage of COSMOS (open diamonds) as well as the counts from \citet[, asterisks]{Papovich04} and \citet[, open squares]{Chary04} are shown for comparison. Our error bars only take into account the statistical Poisson uncertainties. Our selection outside the masked regions is mostly complete for sources at 80\muJypa\,$\ltapp$\,S$_{\rm 24\mu m}$\,$\ltapp$\,3\,mJy. The dotted line shows our extrapolation of the COSMOS source number counts at fluxes below 80\muJy (see Sect.\,\ref{sec:background}). [{\it See the electronic edition of the Journal for a color version of this figure.}]}
\label{fig:counts}
\end{figure}

The differential 24\mic source number counts normalized to the euclidean slope and derived from our COSMOS data are shown in Figure~\ref{fig:counts}. They are compared with the counts obtained by \citet{Papovich04} based on the MIPS {\it Guaranteed Time Observer} programs and by \citet{Chary04} using deep observations in the ELAIS-N1 field.  Within the uncertainties our results show a quite good agreement with these two other surveys, and the variations between the three determinations may result from the effect of cosmic variance between the different fields observed with \spi. At fluxes brighter than $\sim$\,3\,mJy though, the catalog that we derived outside of the optically-masked regions of COSMOS underestimates the density of sources compared to the counts derived over the whole area of the field. This is not surprising given that many of the 24\mic bright objects are associated with optically-bright counterparts in these masked regions.  Such objects are either stars or very low redshift galaxies, which should not affect too much our analysis of the general extragalactic 24\mic source population.

Similar to what was observed in other {\it Spitzer } 24\mic deep surveys \citep{Marleau04,Papovich04,Chary04}, the counts measured from our COSMOS data show a prominent break at flux densities $S_{\rm 24\mu m}$\,$\sim$\,0.2--0.3\,mJy. Our source extraction is highly reliable at these flux levels and this feature can not be artificially produced by an incompleteness of our 24\mic catalog.  Such a rapid ``turn-over" in the euclidean-normalized source number counts implies their rapid convergence at fainter fluxes ($S_{\rm 24\mu m}$\,$\ltapp$\,0.1\,mJy).  Hence, a large fraction of the 24\mic background must be produced by sources readily detected in our COSMOS MIPS data. For instance \citet{Papovich04} assumed a straight extrapolation of the faint-end slope of the counts measured down to $S_{\rm 24\mu m}$\,$\sim$\,60\muJy and they argued that MIPS-24\mic sources detected above 60\muJy resolve as much as $\sim$\,70\% of the 24\mic background intensity.  In Sect.\,\ref{sec:background} we will use our photometric redshift identifications in COSMOS to explore the build-up of this mid-IR background light {\it as a function of cosmic time}.

Finally, we note that the persistent ``shoulder" appearing at $S_{\rm 24\mu m}$\,$\sim$\,3\,mJy in the normalized counts that were derived from large area surveys such as the {\it Spitzer Wide-Area Infrared Extragalactic Survey\,} \citep[SWIRE,][]{Shupe08} is also visible in COSMOS (although our data alone can not formally exclude an effect of cosmic variance). To our knowledge this characteristics of the 24\mic counts has not been explained so far.

\vspace{.5cm}

\section{Observed redshift distributions}
\label{sec:z_distrib}

\begin{figure*}[htpb]
  \epsscale{1.1}
  \plotone{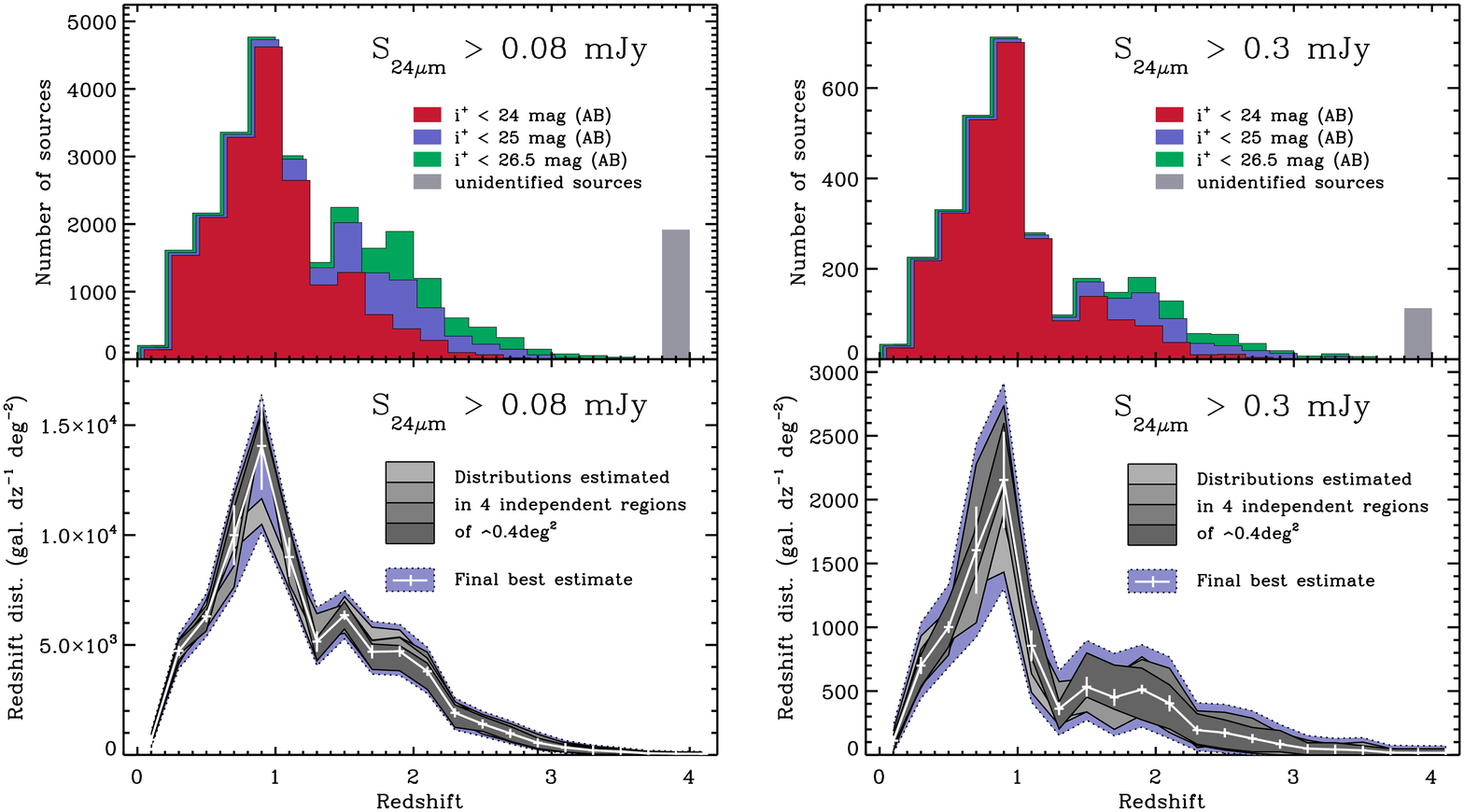}
  \caption{Redshift distributions $\mathcal H_{\rm 24} (z, i^+_{\rm lim}, S^{\rm 24\mu m}_{\rm lim})$ of 24\mic sources with $S^{\rm 24\mu m}_{\rm lim}$\,=\,80\muJy (left) and $S^{\rm 24\mu m}_{\rm lim}$\,=\,300\muJy (right). The top panels represent the redshift histograms determined for different cuts of $i^+_{\rm AB}$--band magnitude as indicated by each legend.  The contribution of objects with no redshift ($i^+_{\rm AB}$\,$>$\,26.5\,mag) is represented as a single bin on the right hand side of the diagrams.  The bottom panels show our final and best estimates of $\mathcal H_{\rm 24} (z, S^{\rm 24\mu m}_{\rm lim})$ along with their corresponding 1$\sigma$ uncertainties (white solid lines). They were derived from the range of possible redshift distributions defined by the dotted lines and obtained after accounting for the effect of cosmic variance and Poissonian uncertainties. The variance produced by large scale structures was quantified by computing the redshift distributions in 4~different $\sim$0.4\,deg$^2$ regions of COSMOS and with Monte-Carlo simulations accounting for the flux and redshift uncertainties of the MIPS sources (grey shaded areas).  Galaxies brighter than $i^+_{\rm AB}$\,=\,20\,mag were excluded from all samples to minimize the cosmic variance fluctuations at low redshift.  [{\it See the electronic edition of the Journal for a color version of this figure.}]}
\label{fig:histo_z}
\end{figure*}

We now explore the redshift distribution of the COSMOS MIPS-24\mic sources based on the identifications of their photometric redshifts described in Sect.\,\ref{sec:photoz}.  We define $\mathcal H_{\rm 24} (z, i^+_{\rm lim}, S^{\rm 24\mu m}_{\rm lim})$ the redshift distribution of the 24\micpa--selected sources with $S_{\rm 24\mu m}$\,$>$\,$S^{\rm 24\mu m}_{\rm lim}$ and associated with optical counterparts at $i^+$\,$<$\,$i^+_{\rm lim}$, and we note $\mathcal H_{\rm opt} (z, i^+_{\rm lim})$ the redshift distribution of all the COSMOS optically--selected sources with $i^+$\,$<$\,$i^+_{\rm lim}$. Given the relatively small number of AGN-dominated sources in the general population of optically-selected galaxies\footnote{Only 0.8\% of optical sources with $i^+$\,$<$\,24\,mag are detected in the $XMM$ observations of COSMOS down to F$_{\rm 0.5-2\,keV}\sim5 \times 10^{-16}$ erg\,cm$^{-2}$\,s$^{-1}$.}, $\mathcal H_{\rm opt} (z, i^+_{\rm lim})$ was computed directly using the photometric redshift catalog of \citet{Ilbert09}. In Figure~\ref{fig:histo_z} we show the redshift distributions $\mathcal H_{\rm 24} (z, i^+_{\rm lim}, S^{\rm 24\mu m}_{\rm lim})$ calculated down to flux limits of 80\muJy and 300\muJy at 24\mic and for $i^+$--band magnitudes limits of 24, 25 and 26.5\,mag. Sources brighter than $i^+_{\rm AB}$\,$=$\,20\,mag were excluded from all these estimates to minimize the effect of large scale structures at low redshift. Our redshift distribution at $z$\,$\ltapp$\,0.4 should actually be treated with caution given the limited comoving volume sampled by our survey at these small distances.

We see that to the depth of our MIPS observations ($S_{\rm 24\mu m}$\,$>$\,80\muJypa) almost all 24\mic sources at $z$\,$\ltapp$\,1 are associated with optical counterparts brighter than $i^+_{\rm AB}$\,$\sim$\,24\,mag. Our photometric redshifts are very accurate at these optical magnitudes. Given the statistics of our sample beyond $z$\,$\sim$\,0.5 our redshift distributions at 0.5\,$\ltapp$\,$z$\,$\ltapp$\,1 should thus be highly robust independently of the 24\mic flux limit considered above 80\muJypa. At $z$\,$\gtapp$\,1, the contribution of fainter sources becomes more significant. For instance, 45\% and 33\% of the MIPS-24\mic sources with respectively $S_{\rm 24\mu m}$\,$>$\,80\muJy and $S_{\rm 24\mu m}$\,$>$\,300\muJy at $z$\,$>$\,1 are associated with counterparts fainter than $i^+_{\rm AB}$\,$=$\,24\,mag. As we showed in Sect.\,\ref{sec:photoz} their photometric redshifts are still quite reliable though (see Figures~\ref{fig:z_uncert} \&~\ref{fig:double_z}).

\begin{figure}
  \epsscale{1.1}
  \plotone{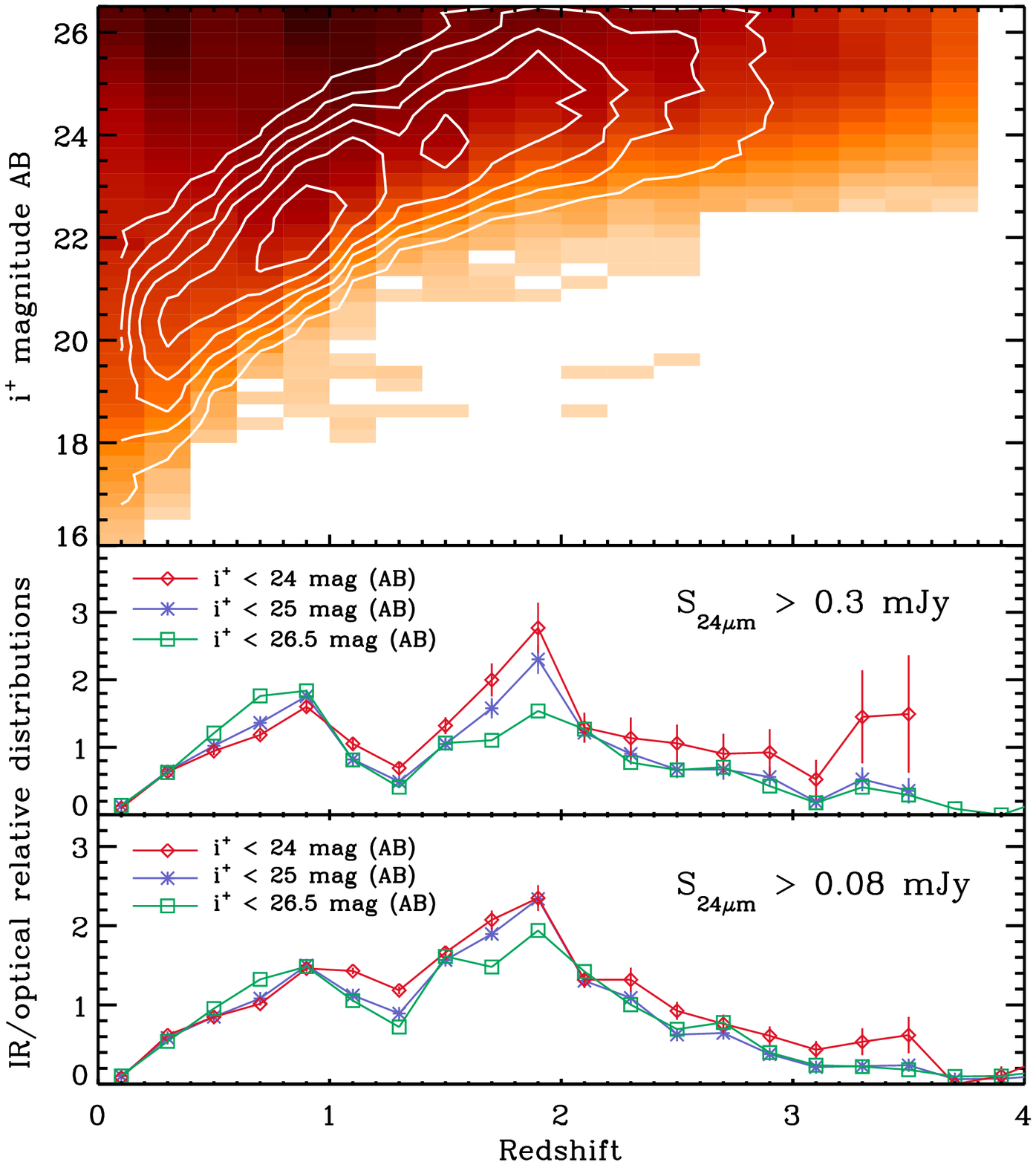}
  \caption{The top panel shows the distribution of $i^+$--band magnitudes as a function of redshift computed for the COSMOS $i^+$--band selected galaxies (Ilbert et al. 2009, shaded region, darkness increasing with source density) and the 24\mic sources with $S_{\rm 24\mu m}$\,$>$\,80\muJy (white contours). The two other panels illustrate the normalized relative distributions $\mathcal H_{\rm 24} (z, i^+_{\rm lim}, S^{\rm 24\mu m}_{\rm lim}) / \mathcal H_{\rm opt} (z, i^+_{\rm lim})$ for $i^+_{\rm lim}$\,=\,24\,mag (open diamonds), $i^+_{\rm lim}$\,=\,25\,mag (asterisks) and $i^+_{\rm lim}$\,=\,26.5\,mag (open squares), with $ \mathcal H_{\rm opt} (z, i^+_{\rm lim})$ the redshift distribution of the optically-selected sources with $i^+_{\rm AB}$\,$<$\,$i^+_{\rm lim}$ derived by \citet{Ilbert09}. They were computed for $S^{\rm 24\mu m}_{\rm lim}$ of 0.3\,mJy and 0.08\,mJy (middle and bottom panels, respectively). The fraction of optical sources detected at 24\mic is clearly rising up to $z$\,$\sim$\,2, although a prominent break is visible at $z$\,$\sim$\,1.3.  [{\it See the electronic edition of the Journal for a color version of this figure.}]}
\label{fig:I_dist}
\end{figure}

Our redshift distribution estimates are mainly affected by the errors on the COSMOS photometric redshifts, the cosmic variance and the errors on the 24\mic flux densities. To quantify the impact of all these uncertainties we first divided the COSMOS field into 4 different regions of similar areas ($\sim$0.4\,deg$^2$) and we derived the possible range of redshift distributions in each region with a set of 5\,000 Monte Carlo simulations. For each of these simulations we considered our {\it full\,} 24\mic catalog down to $S_{\rm 24\mu m}$\,$=$\,60\muJy and we modified the 24\mic flux and redshift of each single source according to their 24\mic flux uncertainty and their 1$\sigma$ photometric redshift dispersion characterized in Sect.\,2. Assuming a given 24\mic flux limit, an average and independent redshift distribution was estimated for the 4 different regions and the dispersion between these 4 estimates was taken as the final uncertainty affecting the global 24\mic source redshift distribution.  Our different estimates are reported in Table\,1. They are also shown in the bottom panels of Figure~\ref{fig:histo_z} for 24\mic flux limits of 80\muJy and 300\muJypa. They suggest that our global redshift distributions inferred from the 1.68\,deg${^2}$ area observed with MIPS should be robust within $\sim$20\% up to $z$\,$\sim$\,3.

It is important to note that the redshift distribution characterizing the galaxy population selected in the mid-IR differs quite significantly from the one observed at optical wavelengths.  Figure~\ref{fig:I_dist} illustrates how the optically and mid-IR selected sources differ in their occupation of the redshift/$i^+$--band magnitude parameter space.  For instance we find that only $\sim$\,27\% of the optically-selected galaxies with $i^+_{\rm AB}$\,$<$\,24\,mag are located at $z$\,$\gtapp$\,1 \citep{Ilbert09} while this fraction rises to more than 50\% for the 24\mic sources with $S_{\rm 24\mu m}$\,$>$\,80\muJypa.  This large proportion of high-redshift mid-IR galaxies is mostly located at 1\,$\ltapp$\,$z$\,$\ltapp$\,2.5 and it has already been seen in other MIPS high redshift surveys \citep[e.g.,][]{Houck05,Yan05,Caputi06a,Papovich07,Desai08}.  It results primarily from the strong evolution that infrared--luminous galaxies have undergone with time \citep[e.g.,][]{Blain99,Franceschini01,Chary01}.  Given the broad-band aromatic features commonly observed at 6\micpa\,$\ltapp$\,$\lambda$\,$\ltapp$\,20\mic in the spectra of star-forming galaxies it may also be due to galaxy $k$-corrections boosting the mid-IR detectability of sources at 1.5\,$\ltapp$\,$z$\,$\ltapp$\,2.5 when these features are redshifted into the MIPS-24\mic bandpass \citep[e.g.,][]{Lagache04}.

\begin{deluxetable*}{lccc}
\tabletypesize{\scriptsize}
\tablenum{1}
\footnotesize
\tablecaption{Redshift distributions of the COSMOS MIPS-24\mic selected sources $^{(a)}$}
\tablehead{
Redshift  & N(z, S$_{\rm 24}$\,$>$\,0.08\,mJy)  &  N(z, S$_{\rm 24}$\,$>$\,0.15\,mJy) &  N(z, S$_{\rm 24}$\,$>$\,0.3\,mJy) }
\startdata
0.0\,$<$\,$z$\,$<$\,0.2 &  679\,$\pm$\,91           &   310\,$\pm$\,63               &  110\,$\pm$\,20        \\        
0.2\,$<$\,$z$\,$<$\,0.4 &  4722\,$\pm$\,203         &   2356\,$\pm$\,64              &  697\,$\pm$\,62        \\        
0.4\,$<$\,$z$\,$<$\,0.6 &  6317\,$\pm$\,275         &   3296\,$\pm$\,71              &  997\,$\pm$\,44        \\        
0.6\,$<$\,$z$\,$<$\,0.8 &  9991\,$\pm$\,1328        &   4964\,$\pm$\,796             &  1604\,$\pm$\,353      \\        
0.8\,$<$\,$z$\,$<$\,1.0 &  14055\,$\pm$\,2001       &   7144\,$\pm$\,1193            &  2148\,$\pm$\,377      \\        
1.0\,$<$\,$z$\,$<$\,1.2 &  8981\,$\pm$\,904         &   3743\,$\pm$\,471             &  865\,$\pm$\,117       \\        
1.2\,$<$\,$z$\,$<$\,1.4 &  5158\,$\pm$\,465         &   1704\,$\pm$\,167             &  367\,$\pm$\,54        \\        
1.4\,$<$\,$z$\,$<$\,1.6 &  6360\,$\pm$\,144         &   2356\,$\pm$\,155             &  531\,$\pm$\,78        \\        
1.6\,$<$\,$z$\,$<$\,1.8 &  4672\,$\pm$\,288         &   1993\,$\pm$\,117             &  452\,$\pm$\,67        \\        
1.8\,$<$\,$z$\,$<$\,2.0 &  4735\,$\pm$\,264         &   2167\,$\pm$\,203             &  497\,$\pm$\,30        \\        
2.0\,$<$\,$z$\,$<$\,2.2 &  3753\,$\pm$\,203         &   1683\,$\pm$\,150             &  411\,$\pm$\,62        \\        
2.2\,$<$\,$z$\,$<$\,2.4 &  1933\,$\pm$\,65          &   815\,$\pm$\,52               &  205\,$\pm$\,13        \\        
2.4\,$<$\,$z$\,$<$\,2.6 &  1423\,$\pm$\,93          &   598\,$\pm$\,51               &  167\,$\pm$\,27        \\        
2.6\,$<$\,$z$\,$<$\,2.8 &  978\,$\pm$\,110          &   406\,$\pm$\,54               &  126\,$\pm$\,31        \\        
2.8\,$<$\,$z$\,$<$\,3.0 &  535\,$\pm$\,89           &   222\,$\pm$\,32               &  84\,$\pm$\,21         \\        
3.0\,$<$\,$z$\,$<$\,3.2 &  335\,$\pm$\,34           &   125\,$\pm$\,8                &  49\,$\pm$\,6          \\        
3.2\,$<$\,$z$\,$<$\,3.4 &  203\,$\pm$\,30           &   103\,$\pm$\,17               &  43\,$\pm$\,5          \\        
3.4\,$<$\,$z$\,$<$\,3.6 &  138\,$\pm$\,7            &   67\,$\pm$\,14                &  36\,$\pm$\,18         \\        
3.6\,$<$\,$z$\,$<$\,3.8 &  73\,$\pm$\,7             &   42\,$\pm$\,8                 &  18\,$\pm$\,7          \\        
3.8\,$<$\,$z$\,$<$\,4.0 &  53\,$\pm$\,8             &   31\,$\pm$\,9                 &  16\,$\pm$\,5          \\        
4.0\,$<$\,$z$\,$<$\,4.2 &   46\,$\pm$\,10           &   34\,$\pm$\,7                 &  16\,$\pm$\,3                      
\enddata
\tablenotetext{(a)}{~The distributions N(z) refer to the number of sources per redshift unit and per square degree. The 1$\sigma$ error bars account for cosmic variance, Poisson noise, flux and redshift uncertainties as described in Sect.\,\ref{sec:z_distrib}. }
\\
\end{deluxetable*}

To better characterize this large contribution of mid-IR selected sources at high redshift, we considered the relative distribution $\mathcal H_{\rm 24} (z, i^+_{\rm lim}, S^{\rm 24\mu m}_{\rm lim}) / \mathcal H_{\rm opt} (z, i^+_{\rm lim})$. First, this quantity allows us to check if any feature of the MIPS-24\mic galaxy redshift distribution does not artificially result from a bias already present in the underlying optically--selected catalog of photometric redshifts. Besides, it can be seen as the fraction of optical sources with $i^+$\,$<$\,$i^+_{\rm lim}$ and with a mid-IR detection above a 24\mic flux limit $S^{\rm 24\mu m}_{\rm lim}$.  Our different estimates of $\mathcal H_{\rm 24} / \mathcal H_{\rm opt}$ are illustrated in the bottom panels of Figure~\ref{fig:I_dist} for the same 24\mic flux and $i^+$--band magnitude limits as the ones considered previously. We see that independently of these limits the fraction of optical sources detected in the mid-IR progressively rises with redshift, which must be related to the well-known increase of star formation density with lookback time. The fraction peaks at $z$\,$\sim$\,2 before gradually declining at higher redshifts, probably because of the decreasing number of 24\mic detections at $z$\,$\gtapp$\,2.

More interestingly though, we observe a substantial drop of $\mathcal H_{\rm 24} / \mathcal H_{\rm opt}$ between $z$\,$\sim$\,1 and $z$\,$\sim$\,1.3 followed by a steep increase from $z$\,$\sim$\,1.3 and $z$\,$\sim$\,2.  This effect is clearly apparent even with a cut at $i^+$$<$\,24\,mag. Therefore it can not originate from systematics affecting our photometric redshifts. It is rather due to the break characterizing the 24\mic source redshift distributions around $z$\,$\sim$\,1.3 and that does not appear in the smooth distributions observed at optical wavelengths (e.g., I09). Note that this break also seems to be more prominent as the 24\mic flux limit increases.  The redshift distribution derived at $S_{\rm 24\mu m}$\,$>$\,300\muJy looks even like a two-component function with a main and narrow peak of sources at $z$\,$\sim$\,1 followed by a secondary but broader peak at 1.5\,$\ltapp$\,$z$\,$\ltapp$\,2.5 (e.g., Figure~\ref{fig:histo_z}, right panel).  This characteristic of the 24\mic redshift distribution could be artificially produced by a relative increase of mid-IR detections at $z$\,$\sim$\,2 due to the rest-frame 7.7\mic aromatic feature enhancing the apparent 24\mic flux of star-forming galaxies when it is redshifted into the MIPS-24\mic bandpass. On the other hand it may also reflect a smaller rate of detections at $z$\,$\sim$\,1.3 given the possible effect of the 9.7\mic silicate absorption. If this break is trully produced by $k$-correction effects, the stronger effect observed at brighter 24\mic flux might also suggest more prominent features in the more luminous mid-IR selected galaxies.

\vspace{0.5cm}
\section{The build-up of the Mid-Infrared  Background across cosmic ages}
\label{sec:background}

\vspace{0.2cm}

The Cosmic Infrared Background (CIB) refers to the extragalactic background light (EBL) detected between $\sim$\,8\mic and $\sim$\,1000\micpa. It can be seen as a fossil record of all the radiations emitted at IR wavelengths along the process of galaxy formation \citep[e.g.,][]{Elbaz02b}.  Its integrated energy power is comparable to that of the background produced by the stellar component of galaxies in the optical/near-IR wavelength range \citep{Dole06}.  Understanding how the different populations of distant sources contributed to the build-up of the CIB as a function of redshift can thus provide significant constraints to our general understanding of the cosmic growth of structures.
    
Using stacking analysis at 70\mic and 160\mic \citet{Dole06} showed that 24\micpa--selected sources with $S_{\rm 24\mu m}$\,$>$\,60\muJy contribute more than 70\% of the CIB.  The large statistics available in the COSMOS field and the high rate ($\sim$\,95\%) of redshift identifications in our 24\mic catalog give us therefore a unique opportunity to resolve the different contributions of galaxies to the build-up of a large fraction of the EBL.  In this current paper we only explore the history of the monochromatic background light at the wavelength of our MIPS observations, using the differential 24\mic source number counts as a function of lookback time.  Resolving the history of mid-IR source number counts already provides tight constraints on the different models of galaxy formation, especially when they correctly reproduce integrated measures like the total counts or the diffuse background emission but differ in their predictions of the flux/redshift distribution of galaxies responsible for these quantities.  This issue will actually be discussed in Sect.\,\ref{sec:compare_model} where we will compare our data with the predictions of different models to quantify their reliability in the description of IR galaxy evolution. We defer to a subsequent publication the analysis of the total background intensity produced by 24\micpa--selected galaxies over the entire spectral range of the CIB.

\begin{figure*}[htpb]
  \epsscale{1.1}
  \plotone{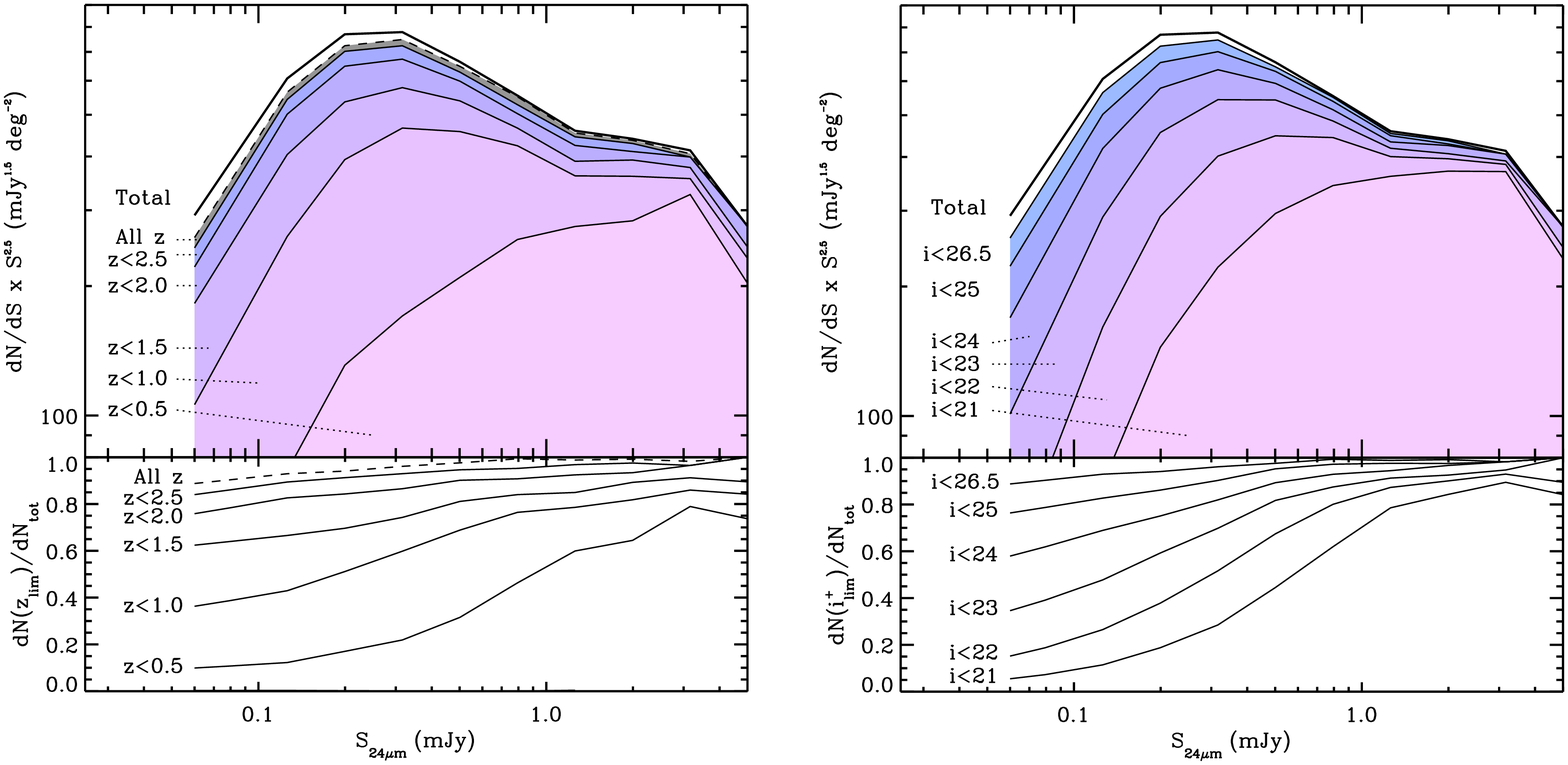}
  \caption{Differential MIPS-24\mic number counts along with the contributions of 24\mic sources computed for different cuts in redshift (left panel) or $i^+$--band magnitude (right panel) and as shown by the shaded regions.  The thick black curve represents the total number counts derived from the whole 24\mic catalog. The bin denoted as ``All z" in the left panel corresponds to all 24\mic sources identified with a redshift in our survey.  The corresponding fractions of sources observed in the different bins of selection are illustrated as a function of 24\mic flux density in the bottom panels.  Note the increasing contribution of optically-faint ($i^+_{\rm AB}$\,$>$\,24\,mag) and high-redshift galaxies to the source number counts at faint 24\mic fluxes.  [{\it See the electronic edition of the Journal for a color version of this figure.}]}
\label{fig:counts_imag}
\end{figure*}

To characterize this decomposition of the 24\mic counts as a function of redshift we first defined $dN(z_{\rm lim},S_{24\mu m})/dS_{24\mu m}$ and $dN(i^+_{\rm lim},S_{24\mu m})/dS_{24\mu m}$ the differential counts of the 24\mic sources identified up to a given redshift~$z_{\rm lim}$ or up to a given $i^+$--band magnitude $i^+_{\rm lim}$, respectively.  In Figure~\ref{fig:counts_imag} we illustrate these quantities computed for $z_{\rm lim}$\,=\,0.5, 1, 1.5, 2 and 2.5 (left panel) and for $i^+_{\rm lim}$\,=\,21, 22, 23, 24, 25 and 26.5\,mag (right panel). For comparison we also show the total number counts derived from the COSMOS field and discussed in Sect.\,\ref{sec:counts} (see also Figure~\ref{fig:counts}).  We see that the brightest 24\mic sources are mostly associated with low redshift galaxies characterized by bright optical counterparts. However a non-negligible fraction of the MIPS detections at $S_{\rm 24\mu m}$\,$\gtapp$\,1\,mJy are also located at much larger distances. Albeit rare in space density, these sources compose the bright end of the IR luminosity function at high redshift and an accurate estimate of their contribution can only be determined with large area surveys.  At fainter fluxes, we also note that a substantial fraction of the 24\mic sources are associated with faint and distant optical counterparts at $i^+_{\rm AB}$\,$>$\,24\,mag. This conclusion reinforces that discussed above for the redshift distributions. It emphasizes once again the difficulties to achieve a complete characterization of the 24\micpa-selected galaxy population with current optical/near-IR follow-ups, and in particular to obtain a full census of the sources at $S_{\rm 24\mu m}$\,$\ltapp$\,0.5\,mJy making the bulk of the mid-IR background light. At such faint optical magnitudes, spectroscopy has been very challenging so far while photometric redshifts also need to be carefully treated given the rapid increase of their uncertainties
 beyond $I$\,$\sim$\,25\,mag (e.g., I09).
 
 To also account for the contribution of galaxies fainter than our 24\mic detection limit we extrapolated the 24\mic number counts at faint fluxes following the same approach as considered by \citet{Papovich04}.  We used power-law functions $dN/dS_{\rm 24 \mu m}$\,$\propto$\,$S_{\rm 24 \mu m}$$^{-\alpha}$ for the full 24\mic catalog and for the different sub-samples of 24\mic sources selected per bin of redshift or $i^+$--band magnitude as described above. For each of these extrapolations the power-law index $\alpha$ was only constrained with the counts observed above $S_{\rm 24\mu m}$\,=\,80\muJy to minimize the impact of the incompleteness of our source extraction below this limit.  For the total counts we derived $\alpha$\,$\sim$\,1.5. This is similar to the results obtained by \citet{Papovich04}, \citet{Chary04} and \citet{Rodighiero06} who measured indexes between 1.5 and 1.6 using sources detected down to $\sim$\,30\muJy in the ELAIS-N1 field.  This good agreement suggests that the extrapolations obtained for the other sub-samples should also provide a fair representation of the flux and redshift distributions characterizing the MIPS sources below our detection limit at 24\micpa.

 Finally, we integrated the total extrapolated number counts to estimate the total 24\mic background light (referred as $B_{\rm tot}$ hereafter). We also integrated the extrapolated functions $dN/dS_{\rm 24 \mu m} \times S_{\rm 24\mu m}$ for each sub-sample considered above to define

\begin{equation}
 B_{24\mu m} (z_{\rm lim}, S_{24\mu m}) = \int_{S_{24\mu m}}^{\infty} dN(z_{\rm lim},S)/dS\times S \times dS
\end{equation} 

~

and 

\begin{equation}
 B_{24\mu m} (i^+_{\rm lim}, S_{24\mu m}) = \int_{S_{24\mu m}}^{\infty} dN(i^+_{\rm lim},S)/dS\times S \times dS
\end{equation}
 
 ~
 
 \noindent as the background light produced above a 24\mic flux $S_{24\mu m}$ and up to the redshift $z_{\rm lim}$ (Eq.\,1) or by sources brighter than the $i^+$--band magnitude $i^+_{\rm lim}$ (Eq.\,2)\footnote{The spectral slope of the counts at faint fluxes guarantees the convergence of the integrals when $S_{24\mu m} \to 0$ whatever the choice of the redshift or the $i^+$--band magnitude limit.}.

\begin{figure}[htpb]
  \epsscale{1.1}
  \plotone{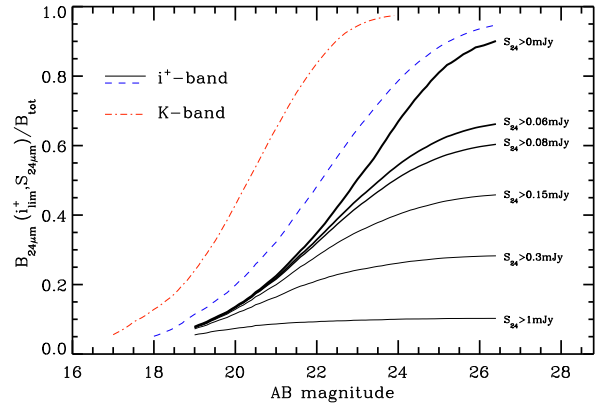}
 \caption{Fraction of the {\it total\,} 24\mic background produced by  sources brighter than the
 $i^+$--band magnitude $i^+_{\rm lim}$
 (black thick solid line, assuming an extrapolation of the counts at the faint end), along with the fractions resolved above 24\mic fluxes of 0.06, 0.08, 0.15, 0.3 and 1\,mJy (solid lines, increasing thickness corresponding to fainter flux limits). For comparison the blue-dashed (resp. red dash-dotted) line shows how the $i^+$--band (resp. $K_s$--band) sources contribute to the fraction of the 24\mic background produced above $S_{\rm 24\mu m}$\,$=$\,80\muJy and directly measured in our data.}
 \label{fig:resolv_imag}
\end{figure}

\begin{figure}[htpb]
  \epsscale{1.1}
  \plotone{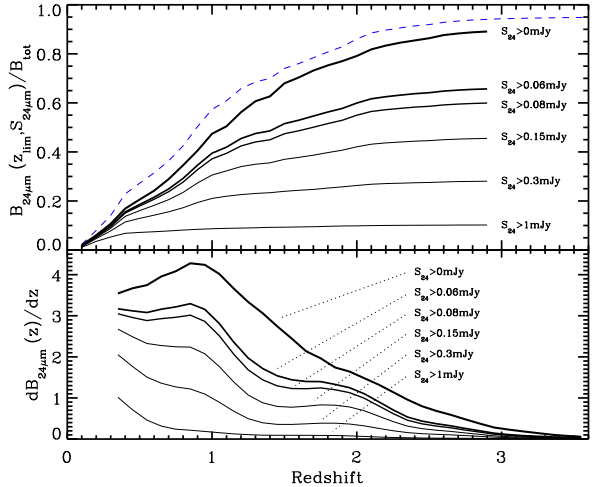}
 \caption{{\it Top:\,} Fraction of the {\it total\,} 24\mic background produced by the COSMOS-selected $i^+$\,$<$\,26.5 mag sources below the redshift $z_{\rm lim}$
 (black thick solid line, assuming  an extrapolation of the counts at the faint end), along with the fractions resolved above 24\mic fluxes of 0.06, 0.08, 0.15, 0.3 and 1\,mJy. For comparison the blue-dashed  line shows how the identified
 24\mic sources contribute to the fraction of the 24\mic background produced above $S_{\rm 24\mu m}$\,$=$\,80\muJy and directly measured in our data.  Uncertainties due to the incompleteness of our redshift identification beyond $z$\,$\sim$\,1 should be at most 10\%.  {\it Bottom:\,} Differential background intensity (arbitrary units) produced as a function of redshift and computed with the same selection criteria as in the top panel.  }
\label{fig:resolv_z}
\end{figure}

 These two quantities are respectively illustrated in Figures~\ref{fig:resolv_imag} \&~\ref{fig:resolv_z} for 24\mic fluxes of 0.06, 0.08, 0.15, 0.3 and 1\,mJy as well as for $S_{24\mu m}$\,=\,0. They were normalized to the total 24\mic background intensity $B_{\rm tot}$ determined from the extrapolation of the total source number counts. They can thus be read as the fraction of the 24\mic background resolved by MIPS-selected sources as a function of their flux $S_{24\mu m}$, their redshift and their optical magnitude.
 The bottom panel of Figure~\ref{fig:resolv_z} also illustrates the {\it differential\,} 24\mic background intensity produced as a function of redshift and computed for the same 24\mic fluxes as in the upper panel.
  Note that because of the magnitude limit imposed by our optical counterpart identifications $B_{24\mu m} (i^+_{\rm lim}, S_{24\mu m})$ was only computed up to $i^+$\,=\,26.5\,mag. Furthermore
  our determination of $B_{24\mu m} (z_{\rm lim}, S_{24\mu m})$ is necessarily underestimated at high redshift given the incompleteness of our identifications at faint fluxes. However we believe that these two effects should affect our results by at most 10\% given our estimate of $B_{24\mu m} (i^+_{\rm lim}, S_{24\mu m})/B_{\rm tot}$\,$\sim$\,$B_{24\mu m} (z_{\rm lim}, S_{24\mu m})/B_{\rm tot}$\,$\sim$\,0.9 for $S_{24\mu m} = 0$, $i^+_{\rm lim} = 26.5$ and $z_{\rm lim} \to \infty$.
 
  While mid-IR shallow surveys (e.g., $S_{24\mu m}$\,$\gtapp$\,0.3\,mJy) only resolve up to $\sim$\,30\% of $B_{\rm tot}$ the analysis of Figures~\ref{fig:resolv_imag} \&~\ref{fig:resolv_z} reveals that the MIPS-24\mic sources brighter than $S_{24\mu m}$\,$\sim$\,60\muJy resolve up to $\sim$\,70\% of the total 24\mic background. This estimate is consistent with that of \citet{Papovich04} but it is slightly higher than the results obtained by \citet{Rodighiero06} in the ELAIS-N1 field. Since a non-negligible fraction of the background is produced by bright sources (e.g., $\gtapp$\,10\% at $S_{24\mu m}$\,$\gtapp$\,1\,mJy) this disagreement could be simply explained by the small area of the sky covered in their analysis ($\sim$\,185\,arcmin$^2$) and the small number of bright 24\mic detections in their data. Furthermore up to $\sim$\,10\%
  of the 24\mic background originates from optical sources fainter than the limit of our photometric redshift catalog ($i^+$\,=\,26.5\,mag), but more than 50\% of $B_{\rm tot}$ is due to sources detected above 80\muJy at $z$\,$\ltapp$\,1.5 and associated with optical counterparts brighter than $i^+$\,=\,24\,mag.  Although their contribution to the integrated CIB would require a careful analysis of their SED in the far-IR, our picture is therefore broadly consistent with previous claims supporting a large contribution of infrared-luminous galaxies at 0.5\,$\ltapp$\,$z$\,$\ltapp$\,1.5 to the EBL \citep[e.g.,][]{Elbaz02b,Dole06}. Conversely, we also find that up to $\sim$\,30\% of the total 24\mic background must be produced by optically-faint ($i^+$\,$>$\,24\,mag) galaxies at $z$\,$\gtapp$\,1. The exact redshift distribution of these sources can not be constrained with our data given our 24\mic flux limit of 80\muJy as well as the incompleteness of our redshift determinations at faint 24\mic fluxes.  However, the flattening of $B_{24\mu m} (z_{\rm lim}, S_{24\mu m})$ observed in Figure~\ref{fig:resolv_z} suggests that they should be mostly located at 1\,$\ltapp$\,$z$\,$\ltapp$\,3.

  Finally, the bottom panel of Figure~\ref{fig:resolv_z} reveals that the relative contribution of galaxies to the 24\mic background increases up to $z$\,$\sim$\,1 and rapidly declines at higher redshift. Although this trend clearly depends on our extrapolations of the counts below 80\muJy it is again consistent with the picture suggested by phenomenological models of IR galaxy evolution \citep[e.g.][]{Chary01,Franceschini01} and by the analysis of the history of the CIB recently performed with other methods \cite[e.g.,][]{Dole06}.  Nonetheless, the contribution of galaxies with $S_{\rm 24\mu m}$\,$>$\,80\muJy seems to be roughly constant up to $z$\,$\sim$\,1. Hence, the ``peak" observed at $z$\,$\sim$\,1 must be produced by a strong evolving (albeit not dominant) contribution of galaxies fainter than 80\muJy and with rather low infrared luminosities.

\vspace{0.5cm}

\section{Comparison with model predictions}
\label{sec:compare_model}

\subsection{Semi-analytical models}
\label{sec:compare_lacey}

Constraining the physics of galaxy formation with semi-analytical models has become over the last decade one of the major challenges of modern cosmology.  Based on the now-popular cold dark matter (CDM) paradigm of structure formation \citep{Springel05} these models compute the evolution of the stellar content of galaxies from the hierarchical assembly of dark matter haloes, using detailed simulations to account for a variety of physical processes such as the radiative cooling and the shock-heating of gas, the energy feedback from supernovae and active nuclei, the chemical enrichment of the interstellar medium and galaxy mergers \citep[e.g.,][]{Cole00,Bower06,Croton06,Cattaneo06}. Although many quantitative aspects of galaxy evolution are still not correctly predicted with these scenarios they have recently been successful in reproducing several key properties of present-day galaxies such as their large-scale clustering or the bimodality of their optical color distribution \citep{Blaizot06,Somerville08}.

Because of the additional difficulty to properly account for dust extinction, the determination of accurate predictions of the deep IR Universe with these models has been however barely explored so far. Simulating the evolution of galaxy IR properties requires not only a correct estimate of the history of the physical processes powering their energy output over relatively short time scales (e.g., star-forming activity, nuclear accretion), but an accurate treatment of their reprocessing of UV/optical radiation at longer wavelengths is also critical. Since the launch of \spi \, \citet{Lacey08} published the only single study (to our knowledge) describing some of the expected properties of 24\micpa-selected galaxies with the use of a semi-analytical model based on the hierarchical growth of dark matter haloes\footnote{\citet{Silva05} derived similar predictions based on the semi-analytical model of \citet{Granato04}, but this model does not include galaxy merging.}. Their simulations were performed with the GALFORM galaxy formation model \citep{Cole00}. They used the GRASIL spectro-photometric code \citep{Silva98} to calculate SEDs and to account for the reprocessing of radiation by dust, which allows for predictions of the multi-wavelength properties of all simulated galaxies as a function of cosmic time.

\begin{figure}[htpb]
  \epsscale{1.1}
  \plotone{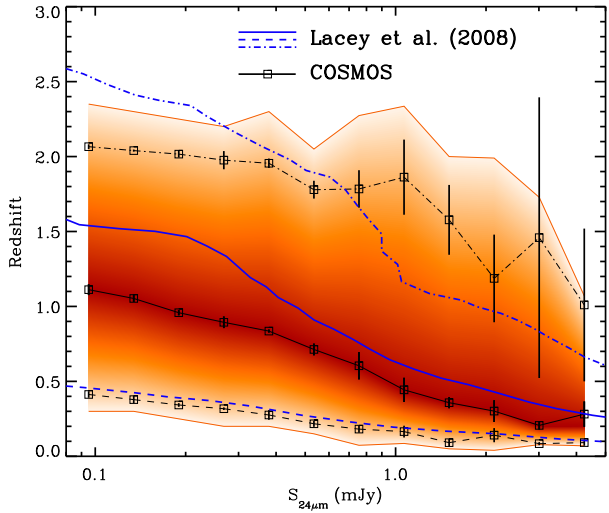}
  \caption{Comparison between our COSMOS flux/redshift distributions (thin lines and open symbols) and the predictions from the semi-analytical model of \citet[][, thick curves, no symbol]{Lacey08}. For each bin of 24\mic flux the solid lines correspond to the median redshift while the dashed and dash-dotted lines represent the 10$^{\rm th}$ and 90$^{\rm th}$ percentiles of the associated distribution. Our error bars account for the uncertainties due to cosmic variance and Poisson noise as described in Sect.\,\ref{sec:z_distrib}.  The shaded region illustrates the parameter space populated by the MIPS-24\mic sources between the 5$^{\rm th}$ and the 95$^{\rm th}$ percentiles of their redshift distribution. This comparison reveals a systematic overestimate of the predicted median redshift at all 24\mic fluxes.  [{\it See the electronic edition of the Journal for a color version of this figure.}]}
  \label{fig:compare_lacey}
\end{figure}

\begin{figure}[htpb]
  \epsscale{1.1}
  \plotone{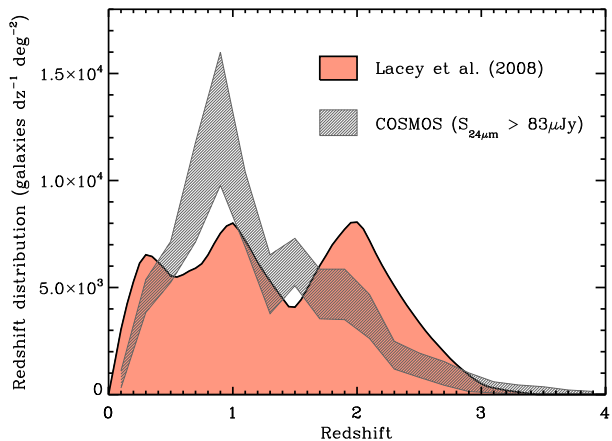}
 \caption{Total redshift distribution of the COSMOS 24\mic sources with $S_{\rm 24\mu m}$\,$>$\,83\muJy  (grey shaded region), compared with the model predictions of \citet[, solid line and filled histogram]{Lacey08}. The observed COSMOS distribution accounts for the uncertainties due to cosmic variance  and Poisson noise as described in Sect.\,\ref{sec:z_distrib}.
   Note the significant excess of sources predicted at $z$\,$\sim$\,2 as well as the lack of 24\mic simulated galaxies at $z$\,$\sim$\,1.}
  \label{fig:compare_z_dist_lacey}
\end{figure}

Figure~\ref{fig:compare_lacey} shows their model predictions for the evolution of the galaxy median redshift as a function of 24\mic flux, along with their estimates of the 10$^{\rm th}$ and 90$^{\rm th}$ percentiles characterizing the associated redshift distributions. They are compared with the same quantities estimated from our COSMOS data, where the reported error bars account for the combination of flux and redshift uncertainties, the cosmic variance and the Poisson noise as described in Sect.\,\ref{sec:z_distrib}. Interestingly enough, we note that the model of \citet{Lacey08} shows a statistically-significant overestimate of the median redshift of galaxies at all fluxes probed by our survey.  This overestimate is particularly noticeable at fluxes below $\sim$\,0.5\,mJy. Given that their predictions agree reasonably well with the total source number counts observed at such flux levels (see their figure~2), it implies an excess of faint sources predicted above $z$\,$\sim$\,1 as well as a lack of sources predicted at lower redshifts.  This disagreement was already noticed by \citet{Lacey08} from the comparison between their model and luminosity functions previously derived from \spi \, data. It is also clearly apparent in Figure~\ref{fig:compare_z_dist_lacey}, where we compare the COSMOS redshift distribution measured above $S_{24\mu m}$\,=\,83\muJy with the result of the simulations that they obtained for the same 24\mic flux limit.  On the other hand we also note that the 90$^{\rm th}$ percentile of the distribution observed in COSMOS at bright fluxes seems to extend up to higher redshifts compared to the model predictions. This could be interpreted as an underestimate of the bright end of the simulated IR luminosity function at 1.5\,$\ltapp$\,$z$\,$\ltapp$\,2.5.  It could also be due to an excess of sources at much lower redshifts since the model of \citet{Lacey08} overestimates the differential 24\mic source number counts above $\sim$\,1\,mJy.  We will investigate these different possible interpretations in future papers where we will present the IR luminosity functions derived from our MIPS imaging of the COSMOS field.

\begin{figure}[htpb]
  \epsscale{1.1}
  \plotone{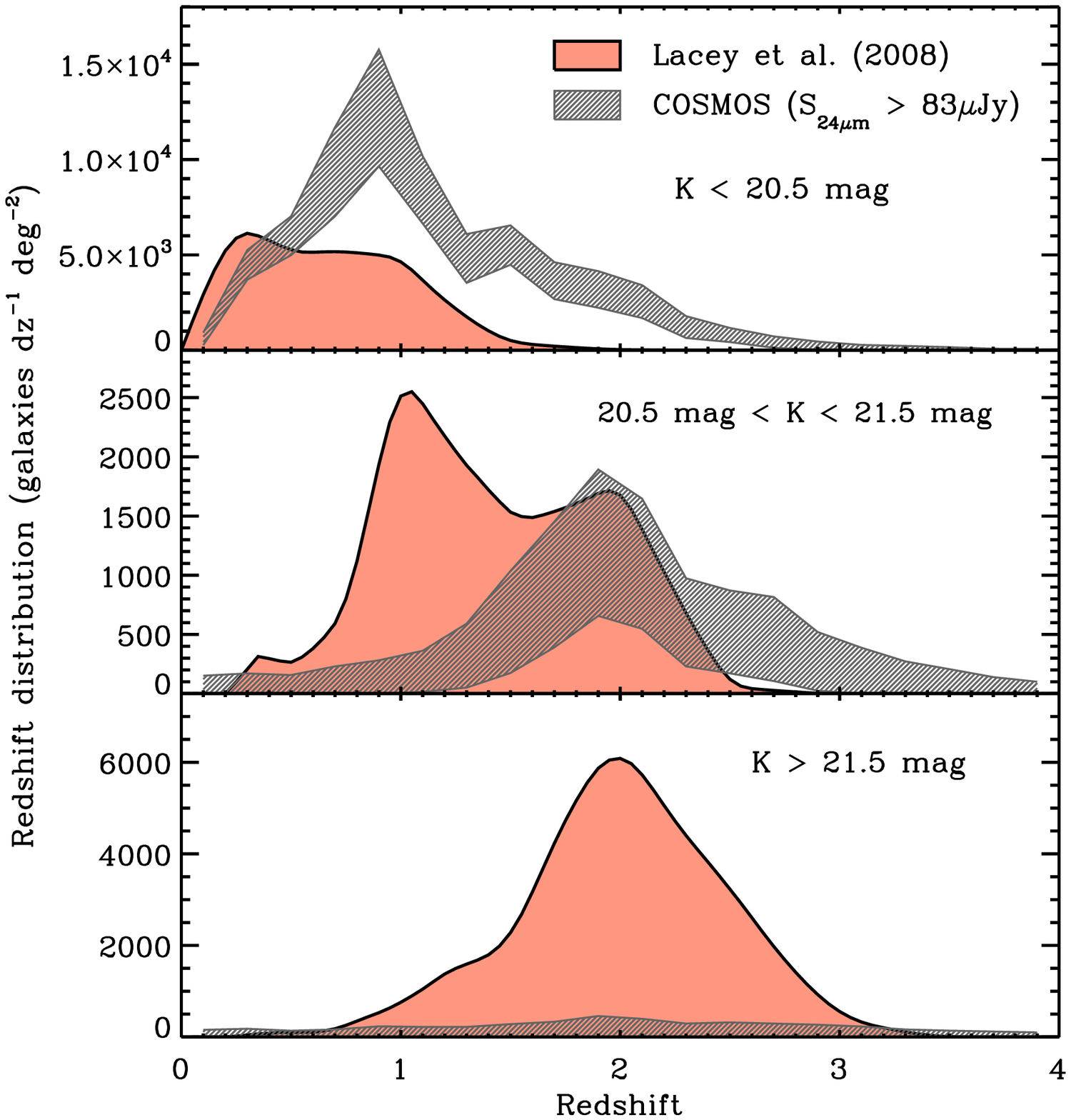}
 \caption{Redshift distributions of  COSMOS 24\mic sources with $S_{\rm 24\mu m}$\,$>$\,83\muJy decomposed into 3~bins of $K_s$--band {\it Vega\,} magnitude (grey shaded regions), compared with the model predictions of \citet[, solid lines and filled histograms]{Lacey08}. The observed distributions account for the uncertainties due to cosmic variance  and Poisson noise as described in Sect.\,\ref{sec:z_distrib}.
   Note the significant excess of sources predicted at $z$\,$\sim$\,2 and mostly due to galaxies fainter than $K_s$\,$\sim$\,21.5\,mag, as well as the lack of 24\mic simulated sources with bright near-IR counterparts.}
\label{fig:compare_z_vs_K_lacey}
\end{figure}

\begin{figure}[htpb]
  \epsscale{1.1}
  \plotone{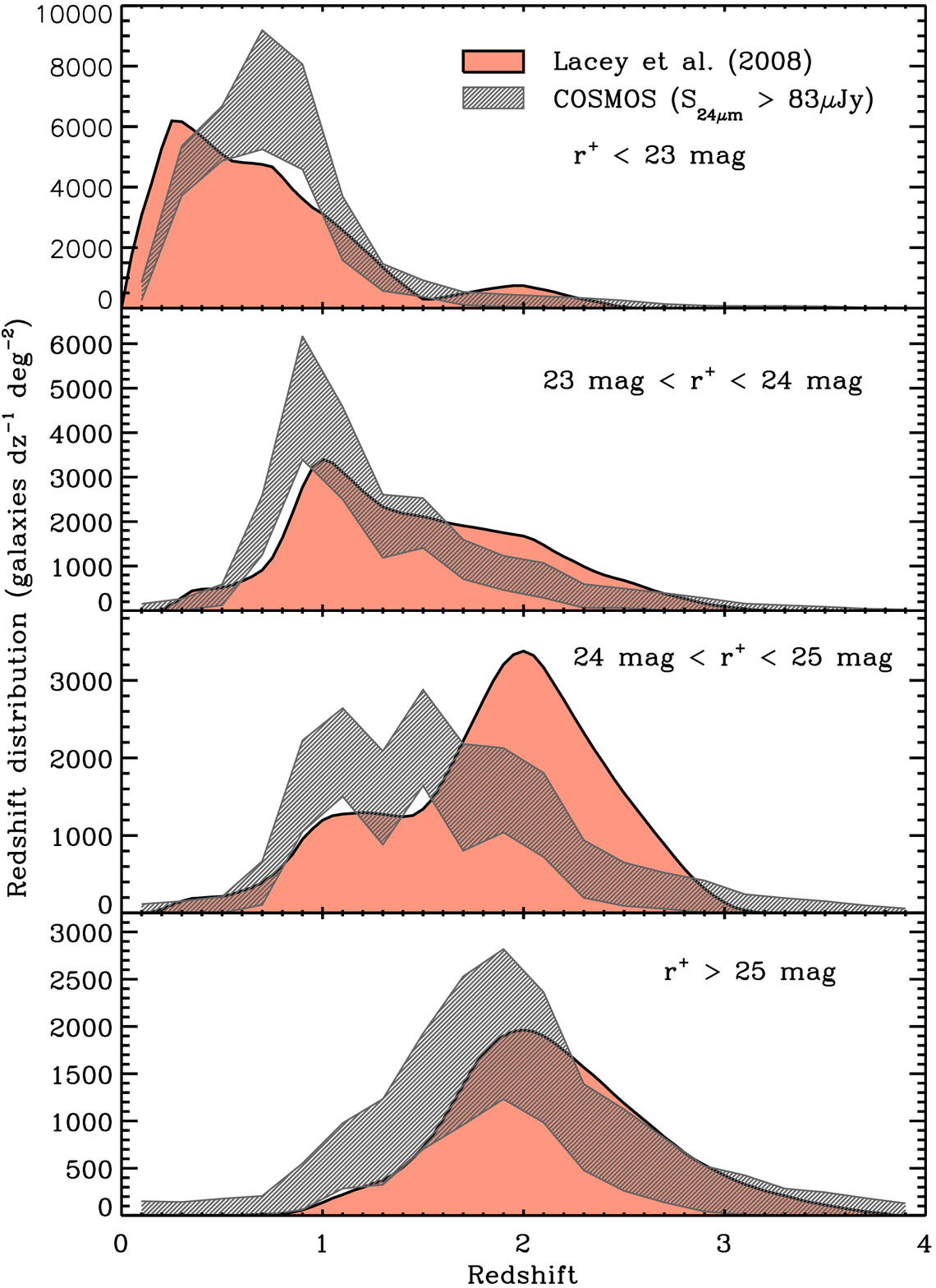}
  \caption{Redshift distributions of COSMOS 24\mic sources with $S_{\rm 24\mu m}$\,$>$\,83\muJy decomposed into 4~different bins of r$^+$--band {\it Vega\,} magnitude (grey shaded regions), compared with the model predictions of \citet[, solid lines and filled histograms]{Lacey08}.  The observed distributions account for the uncertainties due to cosmic variance and Poisson noise as described in Sect.\,\ref{sec:z_distrib}. The predictions show a statistically better agreement than in the $K_s$--band (Fig.\,12), although strong discrepancies can still be noted.}
 \label{fig:compare_z_vs_R_lacey}
\end{figure}

In Figures~\ref{fig:compare_z_vs_K_lacey} and~\ref{fig:compare_z_vs_R_lacey} we now confront the observed and the predicted distributions of 24\mic sources as a function of their near-IR and optical photometry.  Following the same selection criteria as those considered by \citet{Lacey08} we restrict our sample to sources brighter than $S_{\rm 24\mu m}$\,$=$\,83\muJypa. We split these sources in different bins of $K_s$--band and $r^+$--band magnitude and we compare the redshift distributions observed in these bins with the corresponding predictions derived from the simulations\footnote{\citet{Lacey08} provide their predicted distributions using magnitudes in the Vega system. We thus transformed our AB magnitudes assuming $r^+_{\rm Vega}$\,=\,$r^+_{\rm AB} - 0.16$ and $K_{\rm s, Vega}$\,=\,$K_{\rm s, AB} - 1.84$. The transformation between our set of filters ($r^+$ \& $K_s$) and that used by Lacey et al. ($R$ \& $K$) was however neglected.}. The range of distributions shown for the COSMOS data (grey shaded regions) accounts for our best estimate of cosmic variance, flux and redshift uncertainties (see Sect.\,\ref{sec:z_distrib}). The photometric errors affecting the $r^+$--band and $K_s$--band data were not taken into account but their effect should be limited given the adopted bins of magnitudes and the depth of the COSMOS observations ($r^+$\,$<$\,26.8\,mag~AB and $K_s$\,$<$\,23.8\,mag~AB at 5$\sigma$: \citealt{Capak07}, McCracken et al. submitted).

As already noted by \citet{Lacey08}, our figures~\ref{fig:compare_z_vs_K_lacey} \&~\ref{fig:compare_z_vs_R_lacey} show that the excess of simulated sources appearing in their model occurs mostly around $z$\,$\sim$\,2. It originates from galaxies fainter than $K_s$\,$\sim$\,21.5\,mag (AB) while most of the 24\mic sources in our selection ($S_{\rm 24\mu m}$\,$>$\,83\muJypa) are typically brighter than this magnitude limit \citep[see also, e.g.,][]{Papovich07}.  On the other hand our comparison in the $r^+$--band reveals that these simulated sources are not particularly fainter than the typical 24\mic high redshift sources at optical wavelengths. In fact the redshift distribution of the 24\mic sources with faint optical fluxes seems to be reasonably well predicted (although the uncertainties due to the incompleteness of our redshift identification beyond $i_{\rm AB}^+$\,=\,26.5\,mag prevents reaching definite conclusions).  The excess of galaxies in the simulations originates from sources lying among the brightest optical counterparts associated with the 24\mic galaxy population at $z$\,$\sim$\,2 (i.e., $r^+$\,$\ltapp$\,25\,mag Vega) and it may thus be related to a too large number of blue galaxies predicted by the models.

Similarly we find that the lack of simulated galaxies noted by \citet{Lacey08} at $z$\,$\sim$\,1 is related to an underestimate of sources
with relatively bright near-IR counterparts, while the predicted number of galaxies with fainter $K_s$--band luminosities ($K_s$\,$\gtapp$\,20.5\,mag Vega) is clearly too large at this redshift. Given the slightly better agreement found at optical wavelengths, these results reveal again some discrepancies not only in the absolute optical/near-IR magnitudes of the predicted galaxies but also in their colors. We emphasize that this strong lack of 24\mic sources with bright near-IR luminosities (i.e., $K_s$\,$<$\,20.5\,mag Vega) is found at all redshifts from $z$\,$\sim$\,0.5 to $z$\,$\sim$\,2.5. Since the $K_s$--band emission of these distant sources should probe the rest-frame properties of their evolved stellar populations this may reflect a systematic underestimate of the density of massive galaxies with on-going star formation in the distant Universe. Unfortunately though, we do not have access to the mock catalogs where the simulated 24\mic sources were extracted from. More detailed comparisons between the physical properties of the observed and the modeled galaxies will have to be performed to better understand the origin of the discrepancies discussed above.

\subsection{Phenomenological models}
 
As shown above the understanding of the deep IR Universe with semi-analytical simulations still remains an important challenge for extragalactic astronomy. In the past few years, simpler descriptions accounting for the strong evolution of galaxy IR properties and based on ``phenomenological" models have been proposed in the literature. Assuming a backward evolution of the local luminosity function and a library of galaxy spectral energy distributions, these models are usually constrained to reproduce a number of quantities like the Cosmic Infrared Background and/or galaxy IR properties such as the source number counts and redshift distributions. Although they can not be used to follow the history of the physical ingredients driving the evolution of galaxies and the change of their properties with time (e.g., gas accretion, stellar mass build-up, interactions, feedback, ...), they do not require a formalism as complex as the one needed in semi-analytical models. They can also be useful in planing the overall characteristics (e.g., source counts, redshift distribution, ...) of new IR or submillimeter surveys to be performed at yet-unexplored wavelengths.

\begin{figure*}[htpb]
  \epsscale{1.1}
  \plotone{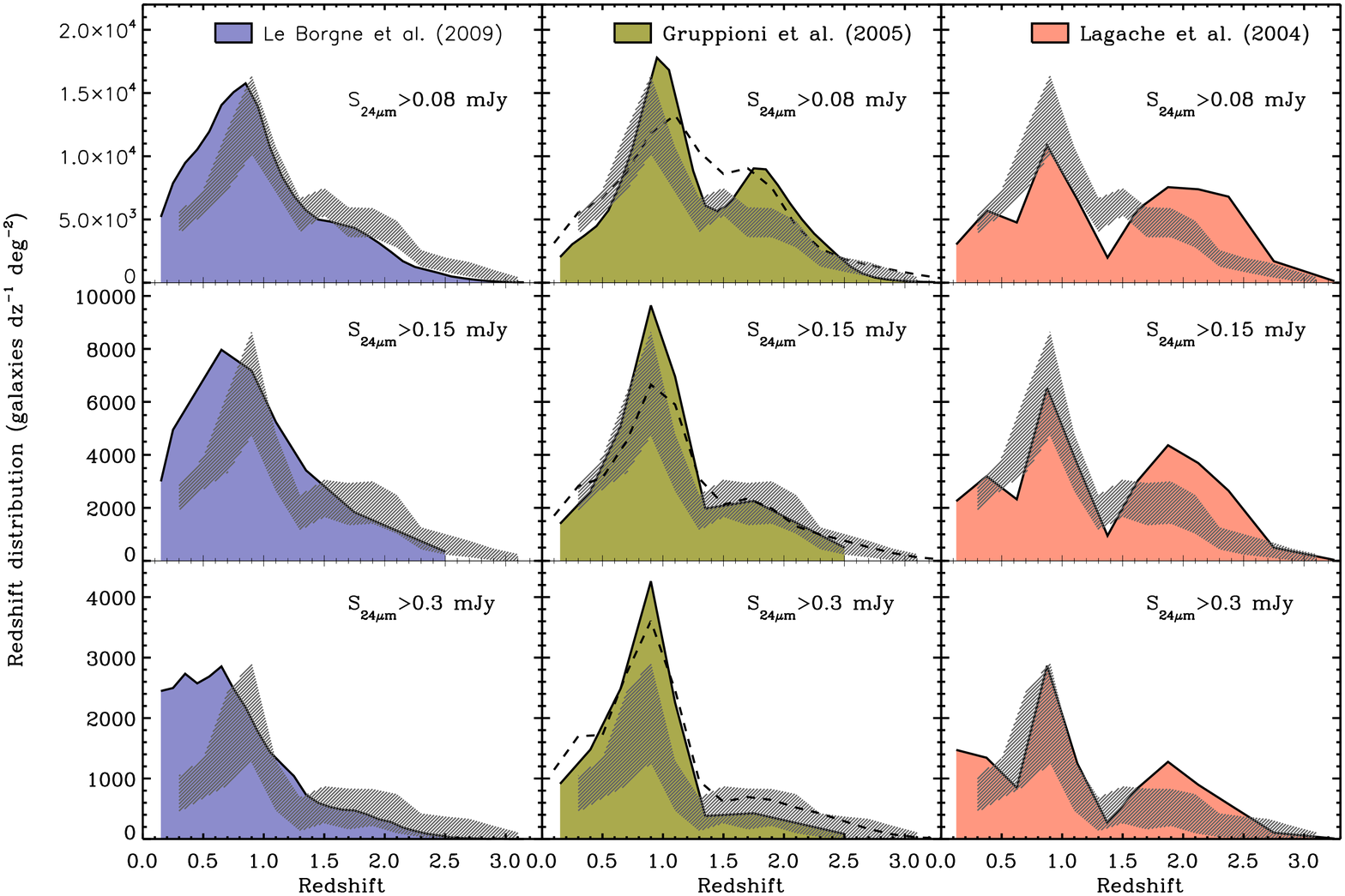}
  \caption{Comparison between our COSMOS redshift distributions (grey shaded regions) and the model predictions (solid lines and filled histograms) of Le~Borgne et al. (2009, left panels), \citet[][, middle panels]{Gruppioni05} and \citet[][, right panels]{Lagache04}, shown for 24\mic flux limits of 0.08, 0.15 and 0.3\,mJy. The observed distributions account for the uncertainties due to cosmic variance and Poisson noise as described in Sect.\,\ref{sec:z_distrib}. The dashed lines in the medium-column panels represent the predictions from a new version of the model developed by Gruppioni et al. (in prep., Gruppioni \& Pozzi, private communication).}
\label{fig:compare_pheno}
\end{figure*}

All these models can vary from one another in the choice of their IR SED library, their decomposition of the overall galaxy population in different sub-groups (e.g., quiescent galaxies, starbursts, active nuclei, ...) or the parameterization characterizing how the luminosity function is allowed to vary with lookback time.  In Figure~\ref{fig:compare_pheno} we present the comparison between our COSMOS 24\mic source redshift distributions and the predictions from the phenomenological models of \citet{LeBorgne09}, \citet[, in prep.]{Gruppioni05} and \citet{Lagache04}. Assuming the luminosity-dependent SED library from \citet{Chary01}, Le~Borgne et al. (2009) have used a non-parametric ``count inversion'' technique with no specific assumption on the evolution of the luminosity function to fit the most-recently updated results from \spi \, IR surveys. On the other hand the model presented by \citet{Gruppioni05} did not make use of any \spi \, data but was constrained with 15\mic observations carried out with the {\it Infrared Space Observatory\,}. It assumes four distinct populations of sources (quiescent galaxies, star-forming sources, type~1 and type~2 AGNs), each of them evolving independently from the others. Initially these populations were characterized by a single template SED \citep{Gruppioni05}, but their scenario was recently updated to account for new procedures fitting individual galaxies with AGN and starburst SEDs simultaneously from the optical up to the mid-IR wavelengths (Gruppioni et al., in prep.).  Finally, \citet{Lagache04} assume a population of luminous starbursts characterized by luminosity-dependent SEDs as well as a population of quiescent cold galaxies evolving independently from the star-forming sources. Their model was updated from a previous version developed before the launch of \spi \, to account for the new constraints imposed at 24\mic by the MIPS cosmological surveys.

The comparisons between these three models and our COSMOS data are illustrated for 24\mic flux limits of 0.08, 0.15 and 0.3\,mJy. The possible range of observed redshift distributions shown in the figure was calculated as described in Sect.\,\ref{sec:z_distrib} by taking into account the combined effect of flux and redshift uncertainties as well as the Poisson noise and the cosmic variance. We did not include any comparison below $z$\,$\sim$\,0.4 as the small comoving volume sampled by our survey at these low redshifts would imply large statistical uncertainties.  At $z$\,$\gtapp$\,1 and faint fluxes (e.g., $S_{\rm 24\mu m}$\,$\ltapp$\,0.15\,mJy) we note that the model of Le~Borgne et al. (2009) shows a rather good agreement with our observations, while the models from \citet{Gruppioni05} and \citet{Lagache04} show a significant overestimate of the number of 24\mic sources around $z$\,$\sim$\,2. As we discussed in Sect.\,\ref{sec:z_distrib} the bimodal distribution predicted by these two models is mainly produced by $k$-correction effects boosting the detectability of starburst sources when their 7.7\mic aromatic feature is redshifted into the MIPS 24\mic bandpass at 1.5\,$\ltapp$\,$z$\,$\ltapp$\,2.5. The predicted bimodality is clearly not as prominent in the data. This disagreement could reflect either the presence of too strong features in the SEDs assumed by the models or a too rapid evolution of the source space density at the faintest fluxes typically probed by our survey.  On the other hand, the break observed at $z$\,$\sim$\,1.3 in the redshift distribution of the COSMOS galaxy population at bright fluxes seems to be better reproduced by the model of Gruppioni et al. than in the predictions from Le~Borgne et al. (where the break is barely apparent) or Lagache et al. (where the break is too prominent due to the strong excess of sources expected at $z$\,$\sim$\,2).  At $z$\,$\sim$\,1 \citet{Gruppioni05} predict however too many sources at $S_{\rm 24\mu m}$\,$\gtapp$\,0.3\,mJy while the model of Le~Borgne et al. may also overestimate the luminosity function at 0.4\,$\ltapp$\,$z$\,$\ltapp$\,0.8.

\vspace{0.5cm}

\section{Discussion}

Because of (i) the unprecedented accuracy of the photometric redshifts used in this work, (ii) our very high rate of redshift identifications obtained down to $S_{\rm 24\mu m}$\,=\,80\muJypa, and (iii) the large area covered by our survey, our analysis represents the most statistically significant characterization of the redshift distributions of the 24\mic source number counts achieved so far. We found that the distribution of the 24\mic selected galaxies is characterized by a prominent peak at $z$\,$\sim$\,1 and a relatively large fraction of sources at 1\,$\ltapp$\,$z$\,$\ltapp$\,3, which is globally consistent with the distributions of MIPS-24\mic sources previously determined with other surveys \citep{Perez-Gonzalez05,Caputi06a,Papovich07,Desai08}.  However, specific differences between our results and these other studies can be noticed. For instance, the redshift distribution obtained by \citet{Perez-Gonzalez05} shows a gradual and rapid decline from $z$\,$\sim$\,1 up to $z$\,$\sim$\,3, while we observe a break at $z$\,$\sim$\,1.3 as well as a larger fraction of sources at $z$\,$\gtapp$\,1. Similarly, our distributions exhibit smoother variations than the redshift distribution determined by \citet{Caputi06a} in the field of the {\it Chandra Deep Field South\,} (CDFS).  All these deviations from our results could originate either from less accurate photometric redshifts that were used in some of these other studies, or from the effect of cosmic variance over small fields of view like the CDFS.

The large number of 24\mic sources at 0.5\,$\ltapp$\,$z$\,$\ltapp$\,1.5 and our analysis of the history of the 24\mic background light also shows that the CIB must originate predominantly from moderately-luminous galaxies at intermediate redshifts rather than highly obscured and luminous dusty sources at $z$\,$\gtapp$\,2. This result has already been infered from previously-published studies \citep[e.g.,][]{Elbaz02b,Dole06} but our data provide for the first time a large enough statistics to explore with good accuracy the build-up of the mid-IR background as a function of cosmic time. The characterization of the far-IR properties of the 24\mic sources with the already-existing MIPS-70/160\mic and MAMBO-1.2\,mm observations of COSMOS \citep[Frayer et al., submitted; ][]{Bertoldi07} or with the future coverage of this field with forthcoming facilities like {\it Herschel\,} will provide an even more detailed picture of the history of the EBL over the whole range of IR wavelengths.

Yet, in spite of this globally coherent picture that has emerged from mid-IR surveys our comparison between the COSMOS 24\mic data and the model predictions of \citet{Lacey08} reveal that the evolution and the properties of luminous galaxies in the distant Universe are still far from being properly reproduced by the semi-analytical simulations of cosmic history.  Although the rapid evolution of the IR luminosity function up to $z$\,$\sim$\,2 is globally well predicted by the model of \citet{Lacey08}, discrepancies in the expected source density per bin of IR luminosity and in the optical/near-IR properties of 24\mic sources are clearly apparent.

Interestingly we found that the number of 24\mic sources with bright near-IR counterparts (i.e., $K_s$\,$\ltapp$\,20.5\,mag Vega) in the simulations is clearly underestimated at all redshifts beyond $z$\,$\sim$\,0.5. It implies that the predicted source density of massive galaxies with strong on-going activity of star formation such as those recently discovered at high redshift with \spi \, \citep[e.g.,][]{Daddi05,Reddy06,Papovich06,Wuyts08} is too low. Besides we note an overestimate of simulated sources with 20.5\,mag\,$<$\,$K_s$\,$<$\,21.5\,mag (Vega) at $z$\,$\sim$\,1, while their counterparts in the $r^+$--band seem to be more representative of the optical properties of 24\mic sources at this redshift.  In addition to the overall underestimate of the predicted source density at this redshift, we thus conclude that a large fraction of simulated galaxies at $z$\,$\sim$\,1 are too blue with respect to the typical 24\mic detected sources. This trend actually recalls the discrepancy that we also observed at $z$\,$\sim$\,2 where the overestimate of predicted sources originates from blue galaxies with $K_s$--band luminosities much fainter than the bulk of 24\mic sources found at this redshift.

The origin of this excess of blue galaxies with a 24\mic detection above our sensitivity limit is not straight forward to determine. Assuming that the star formation rate (SFR) of individual sources is correctly predicted by the simulations, it may be related to an incorrect description of their spectral energy distributions. As we mentioned in Sect.\,\ref{sec:z_distrib} the broad aromatic features commonly observed at 7.7\micpa, 11.3\mic and 12.7\mic in the spectra of starburst galaxies can enhance their detectability when these lines are redshifted into the MIPS-24\mic bandpass.  An overestimate of the strength of the 7.7\mic feature could thus result in too many 24\mic sources predicted at $z$\,$\sim$\,2.  Furthermore the $r^+$--band emission of galaxies at such high redshifts relates to their rest-frame UV properties, which are obviously very sensitive to dust extinction given our selection performed at mid-IR wavelengths.  A wrong treatment of the obscuration affecting these 24\mic sources would thus bias their predicted characteristics at optical wavelengths as well as the estimate of their reprocessed dust emission in the mid-infrared.
 
On the other hand, this excess could also originate from an overestimate of the galaxy star formation rate itself. In the simulations of \citet{Lacey08} galaxies with on-going bursts of star formation account for $\sim$\,50\% and $\gtapp$\,90\% the 24\mic source population at $z$\,$\sim$\,1 and $z$\,$\sim$\,2 respectively (the other sources being classified as quiescent galaxies, see their Figure~A4). Their mid-IR emission directly scales with their star-forming activity, which could lead to a too large number of predicted sources if their SFR is statistically too high. Indeed, star formation processes critically depend on fundamental mechanisms and physical parameters that may still not be fully constrained in the simulations, like for instance the critical density above which stars can form depending on the galaxy internal dynamics and the various energy feedback (AGNs, supernovae, ...).

Finally, it is worth noting that the observed optical and near-IR colors may also depend on the Initial Mass Function (IMF) assumed in the simulations. In the model of \citet{Lacey08} a normal neighborhood IMF leads to a too little evolution of the mid-IR luminosity function compared to the data, and it also underestimates the strong evolution observed in the faint submillimeter number counts \citep{Baugh05}.  A key ingredient characterizing their scenario is the use of a top-heavy IMF for the bursts of star formation triggered by galaxy mergers, which seems to better reproduce the evolution of the galaxy luminosity function observed in the mid-IR. Under this assumption though, the larger fraction of blue and massive stars produced in bursts should lead to bluer stellar populations and this could be one of the reasons explaining the excess of blue galaxies in their predictions. A much more detailed comparison between the intrinsic properties (mass, SFR, SEDs) of {\it individual\,} simulated sources and those characterizing the observed 24\mic galaxy population will be necessary to identify the critical parameters that need to be better constrained for improving model predictions in the infrared.

Given the large number of ingredients to control, the determination of proper predictions of galaxy evolution at IR wavelengths with current semi-analytical models remains obviously a challenging task. Reproducing the evolution of the IR luminosity function requires not only an accurate description of the evolution of the bolometric luminosity function for star-forming galaxies but also an accurate treatment of dust extinction as a function of galaxy properties and a reliable representation of galaxy SEDs in the infrared. Furthermore, dust-obscured star-forming galaxies are not the only contributers to the number counts at IR wavelengths. Quasars and dusty AGNs should also be taken into account, especially at the bright end of the luminosity function where their contribution is far from being negligible \citep[e.g.,][]{Brand06}. Given the coeval growth of bulges and super-massive black holes \citep[e.g.,][]{Page01} ``composite" sources experiencing both nuclear and star-forming activity may actually represent a large fraction of the mid-IR selected galaxy population at high redshift. This large variety of physical properties characterizing IR-luminous galaxies added to the complexity of their spectral signatures in the mid-IR can thus make the predictions from semi-analytical simulations particularly difficult at IR wavelengths.

This contribution of nuclear accretion to cosmic history has not been implemented in the model of \citet{Lacey08}. Therefore the feedback action of AGNs, which has been recently discussed in the framework of other semi-analytical models as an efficient process to regulate the star-forming activity in galaxies, is not taken into account in their simulations. Instead, the decrease of gas cooling for the formation of massive spheroids is controlled in their model by the action of superwinds driven by supernova explosions. These winds produce a feedback effect qualitatively similar to that of AGNs, but the underlying physical mechanisms are somehow different and they result in different predictions for the evolution of the galaxy luminosity function with redshift \citep{Lacey08}. Infrared simulations of distant sources accounting for the role of AGN feedback were actually presented by \citet{Silva05} using the model of \citet{Granato04}. However their scenario did not directly include the merging of galaxies and dark matter haloes, and contrary to our results their 24\mic source redshift distribution at $S_{\rm 24\mu m}$\,$>$\,0.2\,mJy was predicted to peak at $z$\,$\sim$\,2.  Given the current $\Lambda$CDM paradigm as well as the role that nuclear accretion played in the evolution of dusty high redshift galaxies, one possible approach to progress in our understanding of the deep IR Universe with semi-analytical models would be therefore to explore the impact of scenarios including both the hierarchical assembly of dark matter haloes and the contribution of AGNs.

\vspace{0.5cm}

\section{Summary}

We performed our very first analysis of the deep 24\mic observations carried out with the {\it Spitzer Space Telescope\,} over the 2\,deg$^2$ of the COSMOS field. We detected almost $\sim$\,30\,000 sources down to a 24\mic flux of 80\muJypa. Using the multi-wavelength ancillary data available in COSMOS we identified the optical/near-IR counterparts and the photometric redshifts for $\sim$\,95\% of these 24\mic detections. The main results that we derived can be summarized as follows:

\vspace{.2cm}

1. The redshift distribution of the 24\micpa--selected galaxies shows a prominent peak at $z$\,$\sim$\,1 but more than $\sim$\,50\% of the sample is  located at 1\,$\ltapp$\,$z$\,$\ltapp$\,3. Below $z$\,$\sim$\,1 almost all 24\mic sources are brighter than 24\,mag (AB) in the $i^+$--band, while a non-negligible of the MIPS sources at higher redshifts are associated with very faint optical counterparts (i.e., 24\,mag\,$\ltapp$\,$i^+$$\ltapp$\,26.5\,mag).

\vspace{.2cm}

2. Below a fixed optical magnitude limit the fraction of $i^+$--band sources with 24\mic detection strongly increases up to $z$\,$\sim$\,2. This reflects the strong evolution that star-forming galaxies have undergone with lookback time. However this rising trend shows a clear break at $z$\,$\sim$\,1.3, which seems to be more pronounced when higher 24\mic flux selections are considered (e.g., $S_{\rm 24\mu m}$\,$\gtapp$\,0.3\,mJy). This feature is also visible in the redshift distributions of 24\mic sources. It probably originates from $k$-correction effects implied by the presence of the strong aromatic features and/or the Silicate absorption commonly observed in the spectra of luminous dusty galaxies.

\vspace{.2cm}

3.  Using extrapolations of the number counts at faint fluxes we resolved the build-up of the mid-infrared background across cosmic ages. We find that $\sim$\,50\% and $\sim$\,80\% of the 24\mic background intensity originate from galaxies at $z$\,$\ltapp$\,1 and $z$\,$\ltapp$\,2 respectively, with a contribution of faint sources reaching $\sim$\,20$\pm$10\% below our detection limit.  As already suggested by other previously-published studies we thus infer that the Cosmic Infrared Background
 is mostly produced by moderately-luminous dusty galaxies at 0.5\,$\ltapp$\,$z$\,$\ltapp$\,1.5 rather than very luminous and obscured sources at $z$\,$\gtapp$\,2.

\vspace{.2cm}

4. The comparison between our results and semi-analytical predictions reveal substantial discrepancies in the simulated redshift distributions as well as in the $R-K$ colors of the 24\mic source counterparts. In particular we found a strong excess of blue sources predicted at $z$\,$\sim$\,2 and the $K$--band fluxes of simulated galaxies appear to be systematically too low at $z$\,$\gtapp$\,0.5.  This could reflect an underestimate of the density of high redshift massive sources with strong on-going star formation possibly related to fundamental physical parameters and/or processes that are still not fully constrained in the simulations of galaxy formation. It could also be due to an inaccurate treatement of dust extinction in the models, which would imply discrepancies in the predictions of the galaxy SEDs at rest-frame UV, optical and mid-IR wavelengths.

\vspace{.2cm}

5. Comparisons with some of the most recent backward evolution scenarios reproduce reasonably well the flux/redshift distribution of 24\mic sources up to $z$\,$\sim$\,3. Nonetheless none of them is able to accurately match our results at all redshifts over the full range of 24\mic fluxes probed by our survey.

\acknowledgments {\it Acknowledgments:\,} It is a pleasure to acknowledge the contribution from all our colleagues of the COSMOS collaboration. More information on the COSMOS survey is available at http://www.astro.caltech.edu/cosmos.
This work is based on observations made with the {\it Spitzer Space Telescope}, a facility operated by NASA/JPL.  Financial supports were provided by NASA through contracts \#1289085, \#1310136, \#1282612 and \#1298231 issued by the Jet Propulsion Laboratory.  Our research project was also supplemented with phenomenological model predictions from published papers or future publications and that Damien Le~Borgne, Carlotta Gruppioni, Francesca Pozzi, Guilaine Lagache and Herv\'e Dole have been willing to share with us; they are greatly acknowledged for their contribution.  We are grateful to David Elbaz and Casey Papovich for insightful comments on our results, and we also appreciated the hospitality of the Aspen Center for Physics where part of our work was prepared. We finally acknowledge the support and the contribution of the {\it Spitzer Science Center\,} to our research program and we warmly thank the referee for a careful reading of the manuscript.  Support for ELF's work was provided by NASA/Caltech through the {\it Spitzer Space Telescope\,} Fellowship Program~(\#1080367). Part of this work was also supported by the grant ASI/COFIS/WP3110 I/026/07/0.

\end{document}